\documentclass[11pt]{iopart}
\usepackage[english]{babel}
\pdfoutput=1
\usepackage{wasysym}
\usepackage{booktabs}
\usepackage{amssymb}
\usepackage{amsbsy}
\usepackage{verbatim}
\usepackage{graphicx}
\usepackage{epstopdf}
\usepackage{color,soul}
\usepackage{bm}
\usepackage{tikz}

\usepackage[colorlinks=true,linkcolor=blue, citecolor=blue, urlcolor=blue, bookmarks]{hyperref}

\usepackage{cite}

\def\bea{\begin{eqnarray}}
\def\eea{\end{eqnarray}}

\def\be{\begin{equation}}
\def\ee{\end{equation}}

\begin{document}

\title{R\'enyi entropies after releasing the N\'eel state in the XXZ spin-chain} 

\author{Vincenzo Alba and Pasquale Calabrese}
\address{International School for Advanced Studies (SISSA),
Via Bonomea 265, 34136, Trieste, Italy, 
INFN, Sezione di Trieste}

\date{\today}

\begin{abstract} 

We study the R\'enyi entropies in the spin-$1/2$ anisotropic Heisenberg chain after a quantum quench starting from the N\'eel state.
The quench action method allows us to obtain the stationary R\'enyi entropies for arbitrary values of the index $\alpha$ 
as generalised free energies evaluated over a calculable thermodynamic macrostate depending on $\alpha$. 
We work out this macrostate for several values of $\alpha$ and of the anisotropy $\Delta$ by solving the thermodynamic Bethe 
ansatz equations. 
By varying $\alpha$ different regions of the Hamiltonian spectrum  are accessed.
The two extremes are $\alpha\to\infty$ for which the thermodynamic macrostate is either the ground state or
a low-lying excited state (depending on $\Delta$) and $\alpha=0$ when the macrostate is the infinite temperature state.
The R\'enyi entropies are easily obtained from the macrostate as function of $\alpha$ and a few interesting limits are analytically characterised. 
We provide robust numerical evidence to confirm our results using exact diagonalisation and a stochastic numerical implementation 
of Bethe ansatz. 
Finally, using tDMRG we calculate the time evolution of the R\'enyi entanglement entropies. 
For large subsystems and for any $\alpha$, their density turns out to be compatible with that of the thermodynamic R\'enyi entropies. 

\end{abstract}

\maketitle

\section{Introduction}

The extraordinary progress in the field of ultracold atomic gases provided the unprecedented opportunity to observe experimentally the 
real time dynamics of isolated many-body quantum systems~\cite{kinoshita-2006,hofferberth-2007,trotzky-2012,
gring-2012,cheneau-2012,langen-2013,fukuhara-2013,langen-2015,pret,kaufman-2016}. 
The theoretical laboratory for studying this fascinating physical phenomenon is the quantum quench, 
in which a system is prepared in a pure state $|\Psi_0\rangle$, and it is let evolve unitarily under the action of a many-body Hamiltonian $H$.
Relevant questions that have been addressed so far include what is the nature of the steady state arising 
at infinite time and whether it is possible to describe it using the paradigm of 
thermalization~\cite{v-29,eth0,deutsch-1991,srednicki-1994,rdo-08,rigol-2012,dalessio-2015}. 
It is nowadays well established that integrable  systems fail to thermalise, contrary to non-integrable ones, because of 
the presence of  relevant conservation laws constraining their dynamics at any time. 
It has been recognised that local properties in the steady state are 
described by a Generalised Gibbs Ensemble (GGE)~\cite{rigol-2007,cazalilla-2006,
barthel-2008,cramer-2008,cramer-2010,fagotti-2008,calabrese-2011,cazalilla-2012a,
calabrese-2012,mc-12,sotiriadis-2012,collura-2013,fagotti-2013,p-13,fe-13b,fcec-14,
kcc14,sotiriadis-2014,ilievski-2015a,alba-2015,essler-2015,cardy-2015,langen-15,
sotiriadis-2016,bastianello-2016,vernier-2016,vidmar-2016,gogolin-2015,
calabrese-2016,ef-16,pe-16,pvw-17}, which is obtained by supplementing the Gibbs 
ensemble with the additional local and quasi-local conserved 
quantities of the model~\cite{ilievski-2016}. 

However, unitarity of the time evolution in quantum mechanics implies that a quenched system can never relax as a whole to 
a statistical ensemble with non zero entropy.
Thus equilibration and thermalisation must be intended at the level of subsystems. 
Given a finite compact subsystem $A$ embedded in an infinite system, its time dependent reduced density matrix is 
$\rho_A(t)\equiv {\rm Tr}_{\bar A} |\Psi(t)\rangle\langle\Psi(t)|$, where the trace is over $\bar A$, the complement of $A$.
The reduced density matrix $\rho_A(t)$ generically corresponds to a mixed state with non-zero entropy which is known as entanglement 
entropy \cite{e-rev}.
The reduced density matrix can have a well defined infinite time limit $\rho_A(\infty)$ with non-zero entropy density. 
A stationary state  is described by a statistical ensemble with density matrix $\rho_E$ for the entire system,
if its reduced density matrix $\rho_{A, E}= {\rm Tr}_{\bar A}(\rho_E)$ equals $\rho_A(\infty)$ 
\cite{barthel-2008,cramer-2008,calabrese-2012,fe-13b,ef-16}.
According to this logic, it is natural that the extensive thermodynamic entropy of the statistical ensemble is 
 nothing but the entanglement accumulated in time\footnote{
In this paper, two entropies are the same when they have the same extensive behaviour, i.e. when their densities 
are  equal. They can have different subleading terms, and in most cases they do.}.

The entropy of a (reduced) density matrix $\rho$ is traditionally measured by the von Neumann form
\be
S[\rho]\equiv -{\rm Tr}\rho\ln \rho\,, 
\ee
but recently alternative measures like the R\'enyi entropies 
\begin{equation}
\label{sr}
S^{(\alpha)}[\rho]\equiv\frac{1}{1-\alpha}\ln\textrm{Tr}\rho^\alpha,
\end{equation}
are becoming more and more popular.
In the limit $\alpha\to1$ one has $S^{(\alpha)}\to S$, but the knowledge of the R\'enyi entropies for different values of the index $\alpha$
gives access to much more information than the von Neumann entropy alone, as for example the entire spectrum of the density 
matrix (see e.g. \cite{calabrese-2008}). 
Furthermore, R\'enyi entropies for integer $\alpha$ are the essence of the replica approach to the entanglement entropy \cite{cc-04}. 
While the replica method was introduced mainly as a theoretical analytic tool to deal with the complexity of $\rho_A$\cite{cc-04}, it became 
a fundamental idea to access the entanglement entropy in stochastic numerical simulations based on Monte Carlo \cite{mc}
and in experiments: R\'enyi entanglement entropies (for $\alpha=2$) have been measured 
experimentally with cold atoms, both in equilibrium~\cite{islam-2015} and after a quantum quench~\cite{kaufman-2016}.

The equivalence between stationary entanglement entropy $S_A(\infty)\equiv S[\rho_A(\infty)]$ (of a finite subsystem $A$ of volume $V_A$
embedded in an infinite system) and thermodynamic entropy $S_E\equiv S[\rho_E]$ (of a large system of volume $V$)
implies that 
\be
\lim_{V_A\to\infty} \frac{S_A(\infty)}{V_A}= \lim_{V\to\infty} \frac{S_E}V\,.
\label{equiv}
\ee
This equivalence 
has been very recently exploited to give an analytic exact prediction for the entire time dependence of the entanglement entropy in integrable 
models \cite{alba-2016} as also carefully tested against numerical simulations in the XXZ spin-chain \cite{alba-2016}.  
Obviously Eq. \eref{equiv} is valid also for the R\'enyi entropies \eref{sr} and it is natural to wonder if and how the results 
of \cite{alba-2016} generalise to these entropies. 

At the same time, another interesting entropy for a non equilibrium quantum system is the diagonal entropy $S_d$~\cite{polkovnikov-2011}, 
which is the von Neumann entropy of the diagonal ensemble with density matrix 
\begin{equation}
\label{rho-diag}
\rho_{d}=\sum\limits_{m=1}^\infty w_m|m\rangle\langle m|\quad\textrm{with}\quad 
w_m\equiv |\langle\Psi_0|m\rangle|^2, 
\end{equation}
where $|m\rangle$ denotes a generic eigenstate of $H$ and $w_m$ its overlap 
with the initial state. In terms of~\eref{rho-diag}, the diagonal R\'enyi entropies 
$S_d^{(\alpha)}$ are 
\begin{equation}
\label{d-renyi}
S_d^{(\alpha)}\equiv\frac{1}{1-\alpha}\ln
\textrm{Tr}\rho_{d}^\alpha=\frac{1}{1-\alpha}\ln\sum\limits_m w_m^\alpha, 
\end{equation}
and of course the diagonal entropy $S_d=-\textrm{Tr}\rho_d\ln\rho_d$ 
can be obtained in the limit $\alpha\to1$. 
The diagonal entropies have the great advantage that can be very easily calculated even for finite systems 
and without the need of solving the many-body dynamics. 
For this reason, a lot of effort has been devoted to understand the relation between the diagonal entropy and the thermodynamic one, i.e.
the stationary subsystem entanglement entropy, see e.g.  \cite{santos-2011,spr-12,gurarie-2013,fagotti-2013b,dls-13,dora-2014,collura-2014,kormos-2014,bam-15,piroli-2016}.
It has been suggested that for integrable models, the diagonal von Neumann entropy is half the thermodynamic entropy \cite{spr-12,gurarie-2013},
a relation that has been proved only recently \cite{ac-17} for a precise class of initial states. 
Furthermore it has been also shown \cite{ac-17} that, for this specific class of initial states,  
the ratio of entropies is  equal to $1/2$ not only for the von Neumann ones, but in general for R\'enyi entropies of arbitrary order. 
It has been subsequently found \cite{btc-17} that some initial states with peculiar symmetries exist such that the ratio between these 
two entropies is different from $1/2$. 
However,  these states have been explicitly constructed only for non-interacting systems and it is unclear whether the
result generalises to interacting models (see also \cite{delfino-14}).

The main goal of this paper is to continue the investigation of the R\'enyi entropies after a quench initiated in \cite{ac-17}.
A first objective will be to substantiate the general results of \cite{ac-17} with an accurate analytic and numerical analysis of the
R\'enyi entropies for a very specific quench: the time evolution of the anisotropic XXZ spin-chain starting from the N\'eel state.
A second one is to study numerically the entanglement R\'enyi entropies by means of tDMRG to start making a connection 
with the quasiparticle picture \cite{calabrese-2005,cc-16} used for the von Neumann entropy in \cite{alba-2016}.

The technique we will use to compute the R\'enyi entropy is the one introduced in \cite{ac-17}, which is an  adaptation of  the 
Thermodynamic Bethe Ansatz (TBA) approach to quantum quenches (overlap TBA or Quench Action method~\cite{caux-2013,caux-16}). 
This technique provides an analytic machinery to compute the R\'enyi entropies both of the diagonal ensemble and of the GGE. 
Within this approach, in the thermodynamic limit, $S_d^{(\alpha)}$ is given  as a generalised free energy
\begin{equation}
\label{r-res}
S_d^{(\alpha)}=\left.\frac{1}{\alpha-1}\Big(2\alpha{\cal E}-
\frac{1}{2}S_{YY}\Big)\right|_{\pmb\rho^\star_\alpha}.
\end{equation}
Here $S_{YY}$ is the Yang-Yang entropy \cite{yy-69} and ${\cal E}\equiv-2\ln|w_m|$ is the strength of the overlap \eref{rho-diag} between 
the eigenstates of the chain and the initial state. 
This form is valid for a specific class of initial states which have non-zero overlap only with parity invariant Bethe states, 
as most of the quenches solved so far \cite{grd-10,cd-12,pozsgay-14,dwbc-14,bdwc-14,pc-14,fz-16,dkm-16,hst-17,msca-16} 
(see however \cite{btc-17}).
In Ref. \cite{ac-17} it has also been shown that after some algebraic manipulations, the R\'enyi entropies of the GGE 
corresponding to the stationary state can be rewritten as 
\begin{equation}
\label{r-gge}
S_{\rm GGE}^{(\alpha)}=\left.\frac{1}{\alpha-1}\Big(4\alpha{\cal E}- S_{YY}\Big)\right|_{\pmb\rho^\star_\alpha},
\end{equation}
showing that $S_{\rm GGE}^{(\alpha)}=2 S_d^{(\alpha)}$ for generic $\alpha$.
In~\eref{r-res} and \eref{r-gge}, $\pmb \rho^*_\alpha$ identifies a  saddle point eigenstate 
(representative eigenstate or thermodynamic macrostate) of the $XXZ$ chain. 
It turned out that $\pmb \rho^*_\alpha$ in~\eref{r-res} depends on the R\'enyi index $\alpha$, and it is not the macrostate 
describing the local observables and the von Neumann entropy~\cite{alba-2016},  which is recovered only for $\alpha=1$. 
This implies that, for generic $\alpha$, $S_{YY}$ in~\eref{r-res} is not the thermodynamic entropy of the GGE. 
An intriguing consequence of this finding is that the steady state contains information about different regions of the spectrum 
of the $XXZ$ chain, which can be accessed by varying $\alpha$. 
This is similar, in spirit, to the observation of Ref.~\cite{garrison-2015} 
that a single eigenstate of a generic (non-integrable) Hamiltonian at 
finite energy density contains information about the full spectrum of the Hamiltonian.

The paper is organised as follows. Section~\ref{sec:mod-quench} introduces 
the $XXZ$ chain, the quench protocol and the Bethe ansatz solution, focusing on the TBA formalism (subsection~\ref{sec:tba}). 
In section~\ref{sec:tba-renyi} we illustrate the TBA calculation of the R\'enyi diagonal and GGE entropies 
and in Section \ref{sec:an}  we present a few limits that can be worked out analytically. 
Section~\ref{sec:de-sum} reports the results for the R\'enyi entropies for general values of $\alpha$; 
numerical checks are also presented in this section using exact diagonalisation (subsection~\ref{sec:ed}) and a stochastic 
numerical implementation of the Bethe ansatz for finite systems (subsection~\ref{sec:ba-num}). 
Finally, in section~\ref{sec:ent-vs-dia} we numerically evaluate the R\'enyi entanglement entropies and extrapolate to 
infinite subsystem size to confirm that they have the same density as the thermodynamic GGE entropies. 
In the conclusions (section~\ref{sec:concl}) we summarise our findings and discuss some possible future directions. 

\section{Model, quench \& Bethe ansatz solution} 
\label{sec:mod-quench}

We consider quantum quenches in the spin-$1/2$ anisotropic Heisenberg chain ($XXZ$ chain) defined by the Hamiltonian 
\begin{equation}
\label{xxz-ham}
{H}=\sum_{i=1}^L\Big[\frac{1}{2}(S_i^+S^-_{i+1}+S_i^+S_{i+1}^-)+
\Delta\Big(S_i^zS_{i+1}^z-\frac{1}{4}\Big)\Big], 
\end{equation}
where $S_i^\alpha$ are spin-$1/2$ operators, and $\Delta$ is the 
anisotropy. We use periodic boundary conditions, identifying 
sites $1$ and $L+1$ of the chain. We restrict ourselves to $\Delta\ge 1$. 
For any $\Delta$, the $XXZ$ Hamiltonian~\eref{xxz-ham} commutes with the 
total magnetisation $S_T^z\equiv\sum_i S_i^z$. Due to the periodic boundary 
conditions,~\eref{xxz-ham} is invariant under one-site translations 
$S_i^\alpha\to S^\alpha_{i+1}$ , i.e., $[{\mathcal T},H]=0$, with ${\mathcal 
T}$ the translation operator. The $XXZ$ model is also invariant under reflections 
with respect to the center of the chain, i.e., $S_i^\alpha\to S_{L-i+1}^\alpha$, 
implying that $[{\mathcal P},H]=0$, with ${\mathcal P}$ the parity operator. 
As a consequence, in the numerical analysis it is convenient to label the eigenstates of~\eref{xxz-ham} 
as $|s_T^z,k,p\rangle$, with $s_T^z,k,p$ the eigenvalues of $S_T^z,{\mathcal T},
{\mathcal P}$.

In the following, we consider the quench from the N\'eel state $|N\rangle\equiv\left|
\uparrow\downarrow\uparrow\downarrow\cdots\right\rangle=\left|\uparrow\downarrow
\right\rangle^{\otimes L/2}$. To exploit translation invariance, we consider the 
combination 
\begin{equation}
\label{in-state}
|\Psi_0\rangle=\frac{|N\rangle+|\bar N\rangle}{\sqrt{2}}, 
\end{equation}
where $|\bar N\rangle\equiv\left|\downarrow\uparrow\right\rangle^{\otimes L/2}$ denotes 
the anti-N\'eel state. At time $t=0$ the chain is prepared in the state $|\Psi_0\rangle$, 
and the subsequent dynamics is generated by~\eref{xxz-ham}. Crucially, $|\Psi_0\rangle$ 
is invariant under all the symmetries of~\eref{xxz-ham}, i.e., ${\cal S}|\Psi_0\rangle=
|\Psi_0\rangle$ for ${\cal S}=S_T^z,{\mathcal T},{\mathcal P}$. Thus, only the eigenstates 
of~\eref{xxz-ham} with $s_T^z=k=0$ and $p=+1$ can have non zero overlap with the 
state~\eref{in-state}. We anticipate (see subsection~\ref{sec:ed}) that using 
these symmetries in exact (full) diagonalisation allows us to obtain all the eigenstates 
with non-zero N\'eel overlap for a chain with $L\approx 22$ sites.

\subsection{Bethe ansatz solution of the $XXZ$ chain} 
\label{sec:ba}

The $XXZ$ chain is exactly solvable by Bethe ansatz~\cite{taka-book}. In the Bethe ansatz 
solution, the eigenstates of~\eref{xxz-ham} in the sector with $M$ down spins (particles), 
i.e., with fixed total magnetisation $S_T^z=L/2-M$, are in correspondence with a set of $M$ 
rapidities $\lambda_j\in{\mathbb C}$. These are obtained by solving the Bethe equations 
\begin{equation}
\label{be}
\left[\frac{\sin(\lambda_j+i\frac{\eta}{2})}{\sin(\lambda_j-i\frac{\eta}{2})}\right]^L=
-\prod\limits_{k=1}^M\frac{\sin(\lambda_j-\lambda_k+i\eta)}{\sin(\lambda_j-\lambda_k-i
\eta)},
\end{equation}
where $\eta\equiv\textrm{arccosh}(\Delta)$. The corresponding eigenstate energy $E$ is 
given in terms of the rapidities as 
\begin{equation}
E=-\sum\limits_{i=1}^M\frac{\sinh^2\eta}{\cosh\eta-\cos(2\lambda_i)}. 
\end{equation}
In the thermodynamic limit the solutions~\eref{be} form string patterns in the complex plane. 
Here the thermodynamic limit $\displaystyle \lim_{\rm th}$ is taken with the number of particles (flipped spins) $N$ and the length $L$ 
going to infinity at fixed density $N/L$.
The rapidities forming a $n$-string, with $n$ being the string length, can be 
parametrised as \cite{taka-book}
\begin{equation}
\label{str-hyp}
\lambda^j_{n,\gamma}=\lambda_{n,\gamma}+i\frac{\eta}{2}(n+1-2j)+\delta^j_{n,\gamma}, 
\end{equation}
where $j=1,\dots,n$ labels the different string components, $\lambda_{n,\gamma}\in{\mathbb 
R}$ is the ``string center'', and $\delta_{n,\gamma}^j$ are the string deviations. 
For most of the eigenstates of~\eref{xxz-ham}, $\delta_{n,\gamma}^j={\mathcal O}(e^{-L})$, 
allowing one to neglect the string deviations~\cite{taka-book} (string hypothesis). 
Physically, $n$-strings describe bound states of $n$ down spins. The string centres 
$\lambda_{n,\gamma}$ are obtained by solving the Bethe-Gaudin-Takahashi (BGT) 
equations~\cite{taka-book}
\begin{equation}
\label{bgt-eq}
L\theta_n(\lambda_{n,\gamma})=2\pi I_{n,\gamma}+\sum\limits_{(n,\gamma)
\ne(m,\beta)}\Theta_{n,m}(\lambda_{n,\gamma}-
\lambda_{m,\beta}). 
\end{equation}
Here $I_{n,\gamma}\in\frac{1}{2}\mathbb Z$ are the BGT quantum numbers and $\Theta_{n,m}$ the 
scattering phases between different string types
\be\fl
\label{Theta}
\Theta_{n,m}(\lambda)\equiv(1-\delta_{n,m})\theta_{|n-m|}(\lambda)+2\theta_{
|n-m|+2}(\lambda)+\cdots+\theta_{n+m-2}(\lambda)+\theta_{n+m}(\lambda). 
\ee
Each different choice of quantum numbers $I_{n,\gamma}$ gives different sets of solutions of~\eref{bgt-eq}, i.e., a 
different eigenstate of~\eref{xxz-ham}. For $\Delta>1$, one has $\lambda_{n,\gamma}\in[-\pi/2,\pi/2)$. In~\eref{bgt-eq} we 
define $\theta_n(\lambda)\equiv2\arctan[\tan(\lambda)/\tanh(n\eta/2)]$. 
The eigenstate energy $E$ and its total momentum $K$ are obtained by summing over 
the rapidities as 
\bea
\label{eps}
E&=&\sum_{n,\gamma}\epsilon_n(\lambda_{n,\gamma}), \qquad {\rm with}\qquad 
\epsilon_n(\lambda)\equiv-\frac{\sinh(\eta)\sinh(n\eta)}{\cosh(n\eta)-\cos(2\lambda)}, \\
K&=&\sum_{n, \gamma}z_n(\lambda_{n,\gamma}), \qquad {\rm with}\qquad 
\quad z_n(\lambda_{n,\gamma})=\frac{2\pi I_{n,\gamma}}{L}. 
\eea
%

In the following, we will also consider the $XXX$ chain, which is obtained by setting 
$\Delta=1$ in~\eref{xxz-ham}. Bethe ansatz  results for the $XXX$ chain can be 
obtained from those for the $XXZ$ model by taking an appropriate scaling limit. 
One first rewrites the formulas for the $XXZ$ chain in terms of the rescaled 
rapidities $\mu$ defined as 
\begin{equation}
\mu\equiv\frac{\lambda}{\eta}\qquad\textrm{with}\qquad\eta\equiv\textrm{arccosh}\Delta. 
\end{equation}
Since $\eta\to 0$ for $\Delta\to1$, the rescaled rapidities $\mu$ are defined in the 
interval $[-\infty,\infty]$. Also, from~\eref{str-hyp} one has that the spacing 
between string components along the imaginary axis becomes $i/2$. Finally, the limit 
$\eta\to 0$ has to be taken. For instance, using~\eref{eps}, 
for the $XXX$ chain $\epsilon_n$ becomes 
\begin{equation}
\epsilon_n(\mu)=\frac{2n}{4\mu^2+n^2}. 
\end{equation}
%

\subsection{Thermodynamic Bethe ansatz (TBA)} 
\label{sec:tba}

In the thermodynamic limit the solutions of the BGT equations~\eref{bgt-eq} become 
dense on the real axis. Local properties of the system can be extracted from the rapidity 
densities $\rho_n(\lambda)$ (one for each string type), which are formally defined as 
\begin{equation}
\rho_n(\lambda)\equiv\lim_{L\to\infty}\frac{1}{L(\lambda_{n,\gamma+1}-\lambda_{n,\gamma})}. 
\end{equation}
To characterise the thermodynamic state of the system one also needs the densities 
$\rho_n^{(h)}(\lambda)$ of the $n$-string holes, i.e., of the unoccupied string centres and it is also custom~\cite{taka-book} 
to introduce the total densities $\rho_{n}^{(t)}(\lambda)\equiv\rho_n(\lambda)+\rho_n^{(h)}(\lambda)$. 

The $\rho_n^{(h)}(\lambda)$ and $\rho_n(\lambda)$ are related via the thermodynamic version of the BGT equations 
\begin{equation}
\label{tba-eq}
\rho_n^{(h)}(\lambda)+\rho_n(\lambda)=a_n(\lambda)-\sum\limits_{m=1}^\infty(a_{nm}
\star\rho_m)(\lambda), 
\end{equation}
which are obtained from~\eref{bgt-eq} by taking the thermodynamic limit. 
In~\eref{tba-eq} we defined $a_{nm}(\lambda)$ as 
\be\fl
\label{anm}
a_{nm}(\lambda)=(1-\delta_{nm})a_{|n-m|}(\lambda)+2a_{|n-m|}(\lambda)
 +\ldots +2a_{n+m-2}(\lambda)+a_{n+m}(\lambda)\,,
\ee
where 
\begin{equation}
a_n(\lambda)=\frac{1}{\pi} \frac{\sinh\left( n\eta\right)}{\cosh (n
\eta) - \cos( 2 \lambda)}\,.
\end{equation}
The convolution $f\star g$ between two functions  is defined as 
\begin{equation}
\left(f\star g\right)(\lambda)=\int_{-\pi/2}^{\pi/2}{\rm d}\mu f(\lambda-\mu)g(\mu)\,.
\end{equation}
Thus in the thermodynamic limit, the total magnetisation and energy densities become 
\bea
\frac{S_T^z}{L}&=&\sum\limits_{n=1}^\infty n\int_{-\frac{\pi}{2}}^{\frac{\pi}{2}}
d\lambda\rho_n(\lambda),\\
\label{c-quant}
\frac{E}{L}&=&\sum\limits_{n=1}^\infty\int_{-\frac{\pi}{2}}^{\frac{\pi}{2}}d\lambda
\epsilon_n(\lambda) 
\rho_n(\lambda), 
\eea
where $\epsilon_n(\lambda)$ is defined in~\eref{eps}. The set of root densities $\pmb{\rho}
\equiv\{\rho_n\}_{n=1}^\infty$ defines a thermodynamic macrostate, and it allows to obtain 
the expectation values of local or quasi-local observables in the thermodynamic limit. 
A thermodynamic macrostate corresponds to an exponentially large (with $L$) number of 
microscopic eigenstates (microstates), which lead to the same set of rapidity densities 
in the thermodynamic limit. Any of the equivalent microscopic eigenstates can be 
chosen as a finite-size representative of the thermodynamic macrostate. The total number 
of equivalent microstates is $e^{S_{YY}}$, with $S_{YY}$ the Yang-Yang 
entropy~\cite{yy-69} 
\be
\label{y-y}
S_{YY}[\pmb{\rho}]\equiv L\sum_{n=1}^\infty\int_{-\frac{\pi}{2}}^{\frac{\pi}{2}}
d\lambda\Big[\rho_n^{(t)}\ln\rho_n^{(t)} -\rho_n\ln\rho_n-\rho_n^{(h)}\ln\rho_n^{(h)}\Big]. 
\ee
Clearly, $S_{YY}$ is extensive. For systems in thermal equilibrium $S_{YY}$ 
is the thermal entropy.

As we shall see, the TBA equations assume a more compact form in terms of the ratio 
\be
\eta_n(\lambda)\equiv \frac{\rho_n^{(h)}(\lambda)}{\rho_n(\lambda)},
\label{eta-def}
\ee
which we define here for later convenience. 

\section{Overlap TBA for the R\'enyi entropies} 
\label{sec:tba-renyi}

In this section we briefly recall the approach of Ref. \cite{ac-17} to calculate R\'enyi entropies (both for the diagonal ensemble and for the GGE)
and then specialise it to the XXZ spin-chain, in particular for the quench from the N\'eel state.  
The quench action ~\cite{caux-2013,caux-16} provides a calculable and manageable representation of the stationary state (or equivalently 
of the diagonal ensemble), as nowadays explicitly worked out for many integrable models \cite{BeSE14,BePC16,pozsgay-13,wdbf-14,PMWK14,ac-16qa,ppv-17,mbpc-17,dwbc-14,pce-16,Bucc16,pvc-16,npg-17,btc-17}.
Furthermore, it can also be  used  to study the time evolution of local observables as done in a few simple cases \cite{dc-14,dlc-15,pc-17}.
In \cite{ac-17} the Quench Action method has been generalised to obtain the stationary values of  the R\'enyi entropies which 
are non-local quantities and indeed the approach introduces important novelties.

For the diagonal ensemble, the starting point is the thermodynamic limit of the overlaps 
$w_m$ appearing in the diagonal ensemble \eref{rho-diag} which can be written as 
\be
\label{def}
w_m=\exp(-2{\cal E}(m)),\qquad {\rm with}\qquad {\cal E}\equiv-\lim_{\rm th} [{\rm Re}(\ln\langle m|\Psi_0\rangle)].
\ee
In the thermodynamic limit the sum over eigenstates appearing in~\eref{rho-diag} and~\eref{d-renyi} is replaced by a functional integral 
over the rapidity densities as
\begin{equation}
\label{rep}
\sum_m\to\int{\mathcal D}\pmb{\rho}e^{S_{YY}[\pmb{\rho}]}, 
\end{equation}
where the factor $e^{S_{YY}}$, with $S_{YY}$ the Yang-Yang entropy \eref{y-y}, 
takes into account the exponentially large number of microscopic eigenstates leading to the 
same densities. In~\eref{rep}, ${\mathcal D}\pmb{\rho}\equiv\prod_{n=1}^\infty{\mathcal D}
\rho_n(\lambda)$. Using~\eref{rho-diag},~\eref{def}, and~\eref{rep} one obtains 
\begin{equation}
\label{qa}
{\rm Tr}\rho^\alpha_{d}=\int{\mathcal D}\pmb{\rho}e^{-2\alpha {\cal E}[\pmb{\rho}]+\frac{1}{2}S_{YY}[\pmb{\rho}]}. 
\end{equation}
The factor $1/2S_{YY}$ takes into account that only parity-invariant 
eigenstates of~\eref{xxz-ham} have non-zero overlap with the N\'eel state~\cite{wdbf-14}. 
These correspond to solutions of the Bethe equations containing only 
pairs of rapidities with opposite sign, i.e., such that $\{\lambda_j\}_{j=1}^M=\{-\lambda_j\}_{j=1}^M$. 

In a similar fashion, it has been shown in \cite{ac-17} that the R\'enyi GGE entropies can be written as
 \begin{equation}
\label{qa-gge}
\textrm{Tr}\rho^\alpha_{\rm GGE}=\int{\mathcal D}\pmb{\rho}e^{-4\alpha {\cal E}[\pmb{\rho}]+S_{YY}[\pmb{\rho}]}. 
\end{equation}
It is evident that both in \eref{qa} and \eref{qa-gge}, the function ${\cal E}$ acts as driving term replacing the Hamiltonian 
in the standard TBA  at finite temperature.

In the thermodynamic limit, the path integrals \eref{qa} and \eref{qa-gge} are dominated by the saddle point, obtained 
by minimising the exponent. Since the exponent in one case is just the double of the other,  $\pmb \rho^*_\alpha$
is the same in the two cases and can be obtained as solution of the saddle point equation
\be
\Bigg[-4\alpha \frac{\delta {\cal E(\pmb\rho)} }{\delta \pmb\rho}+  \frac{\delta S_{YY}(\pmb \rho) }{\delta \pmb\rho}\Bigg]_{\pmb \rho=\pmb \rho^*_\alpha}=0
\label{spe-gen}
\ee
and hence the R\'enyi entropies are
\be
\label{d-renyi-main}
S_d^{(\alpha)}=\frac{1}{1-\alpha}\Big[-2\alpha{\cal E}(\pmb \rho^*_\alpha)+\frac{1}{2}S_{YY}(\pmb \rho^*_\alpha)\Big]= 
\frac12 S_{\rm GGE}^{(\alpha)}.
\ee
Having established that the GGE entropy is just the double of the diagonal entropy, in the following we will just refer to the latter, 
being clear that, at this point, the former does not provide any further information.

We now are ready to specify the general approach to the quench in the XXZ spin-chain from the N\'eel state. 
The overlaps between the N\'eel state and the Bethe eigenstates have been derived in full generality for finite systems in \cite{bdwc-14}.
The extensive part of the thermodynamic limit has been extracted analytically in~\cite{wdbf-14} and it is given as
\begin{equation}
\label{driving}
{\cal E}=\frac{L}{2}\sum_n\int_0^{\frac{\pi}{2}}d\lambda\rho_n(\lambda) \big[g_n(\lambda)+4n\ln 2\big], 
\end{equation}
where 
\begin{equation}
\label{gn}
g_n=\sum\limits_{l=0}^{n-1}\ln\Big[\frac{s_{n-1-2l}c_{n-1-2l}s_{-n+1+2l}c_{-n+1+2l}}{
t_{n-2l}t_{-n+2l}}\Big],
\end{equation}
with
\bea
 s_n(\lambda)&\equiv&\sin\big(\lambda+i\frac{n\eta}{2}\big),\\
 c_n(\lambda)&\equiv&\cos\big(\lambda+i\frac{n\eta}{2}\big),\\
 t_n(\lambda)&=&\frac{s_n(\lambda)}{c_n(\lambda)}.
\eea

Thus, the saddle point equation \eref{spe-gen}  yields the generalised TBA equations 
\be
\label{qa-tba}
\ln[\eta_n(\lambda)]=2n[\alpha\ln4-h]+\alpha g_n(\lambda)
+\sum\limits_{m=1}^\infty a_{nm}\star\ln(1+\eta_m^{-1})(\lambda), 
\ee
which is an infinite system of coupled integral equations for $\eta_n$. 
In~\eref{qa-tba} $a_{nm}$ are the same as in~\eref{tba-eq}. The magnetic field 
$h$ (a.k.a. the chemical potential) has to be introduced to ensure the zero magnetisation condition
\begin{equation}
\label{m-constr}
\sum\limits_{m=1}^\infty m\int_{-\frac{\pi}{2}}^\frac{\pi}{2}
d\lambda\rho_m(\lambda)=\frac{1}{2}. 
\end{equation}
For all the values of $\alpha$ that we considered, we numerically verified that for $\Delta>1$ the constraint~\eref{m-constr} is satisfied. 

Similar to the standard TBA~\cite{taka-book}, it is possible to partially decouple the equations for different $n$ in the system~\eref{qa-tba}. 
This leads to the partially decoupled equations~\cite{wdbf-14} 
\begin{equation}
\label{dec}
\ln\eta_n=\alpha d_n+s\star[\ln(1+\eta_{n-1})+\ln(1+\eta_{n+1})], 
\end{equation}
with $\eta_0=0$ and 
\bea
\label{dn}
 d_n(\lambda)&\equiv& (-1)^n\ln\frac{\vartheta^2_4(\lambda,\tau)}{\vartheta^2_1(\lambda,\tau)}
+\ln\frac{\vartheta^2_2(\lambda,\tau)}{\vartheta^2_3(\lambda,\tau)},\\
 s(\lambda)&\equiv&\frac{1}{2\pi}\sum\limits_{k=-\infty}^\infty \frac{e^{-2 i k\lambda}}{\cosh(k\eta)}. 
\eea
Here $\vartheta_k(\lambda,\tau)$ with $\tau\equiv e^{-2\eta}$ are the Jacobi elliptic theta functions. 
Notice that $d_n(\lambda)$ only depends on the parity of $n$.
A similar decoupling can be obtained for the TBA equations~\eref{tba-eq} for $\rho_n$ as~\cite{taka-book,wdbf-14} 
\begin{equation}
\label{dec-rho}
\rho_n(1+\eta_n)=s\star(\rho_{n-1}\eta_{n-1}+\rho_{n+1}\eta_{n+1}), 
\end{equation}
with the convention  $\rho_0=\delta(\lambda)$ and $\eta_0=1$. 

The solutions $\eta_n$ of~\eref{qa-tba} are used in~\eref{tba-eq} to obtain the saddle point densities $\pmb{\rho}\equiv\{\rho_n\}_{n=1}^\infty$, 
which plugged in~\eref{d-renyi-main} give the R\'enyi entropies. For $\alpha=1$, since $\textrm{Tr}\rho_d=1$, the saddle point is 
\begin{equation}
\label{tri-0}
S_{YY}=4{\cal E}. 
\end{equation}

The coupled integral equations~\eref{qa-tba} and~\eref{tba-eq} admit a unique solution for arbitrary value of $\alpha$.
However, while, for $\alpha=1$ the saddle point densities can be derived analytically~\cite{wdbf-14}, 
for generic $\alpha$ this does not seem to be possible. 
It would be highly desirable to find some new tricks which would allow the solution of these equations. 
Anyhow, in the absence of an analytic solution we should solve these equations numerically. 
Although~\eref{qa-tba} and~\eref{tba-eq} are a set of infinite coupled integral equations, one can truncate the system by including the 
first few strings $n\le n_{\rm max}$, with $n_{\rm max}$ large enough to ensure convergence, as routinely done to solve TBA equations. 
A test of the convergence is provided by monitoring the magnetisation sum rule~\eref{m-constr}. 
We numerically find that, at least for large enough $\Delta$, the convergence with $n_{\rm max}$ is quite fast, and reliable results can be obtained 
using modest values of $n_{\rm max}$. As expected, the value of $n_{\rm max}$ has to be increased when approaching the isotropic limit $\Delta\to 1$. 

%
\begin{figure}[t]
\begin{center}
\includegraphics*[width=0.85\linewidth]{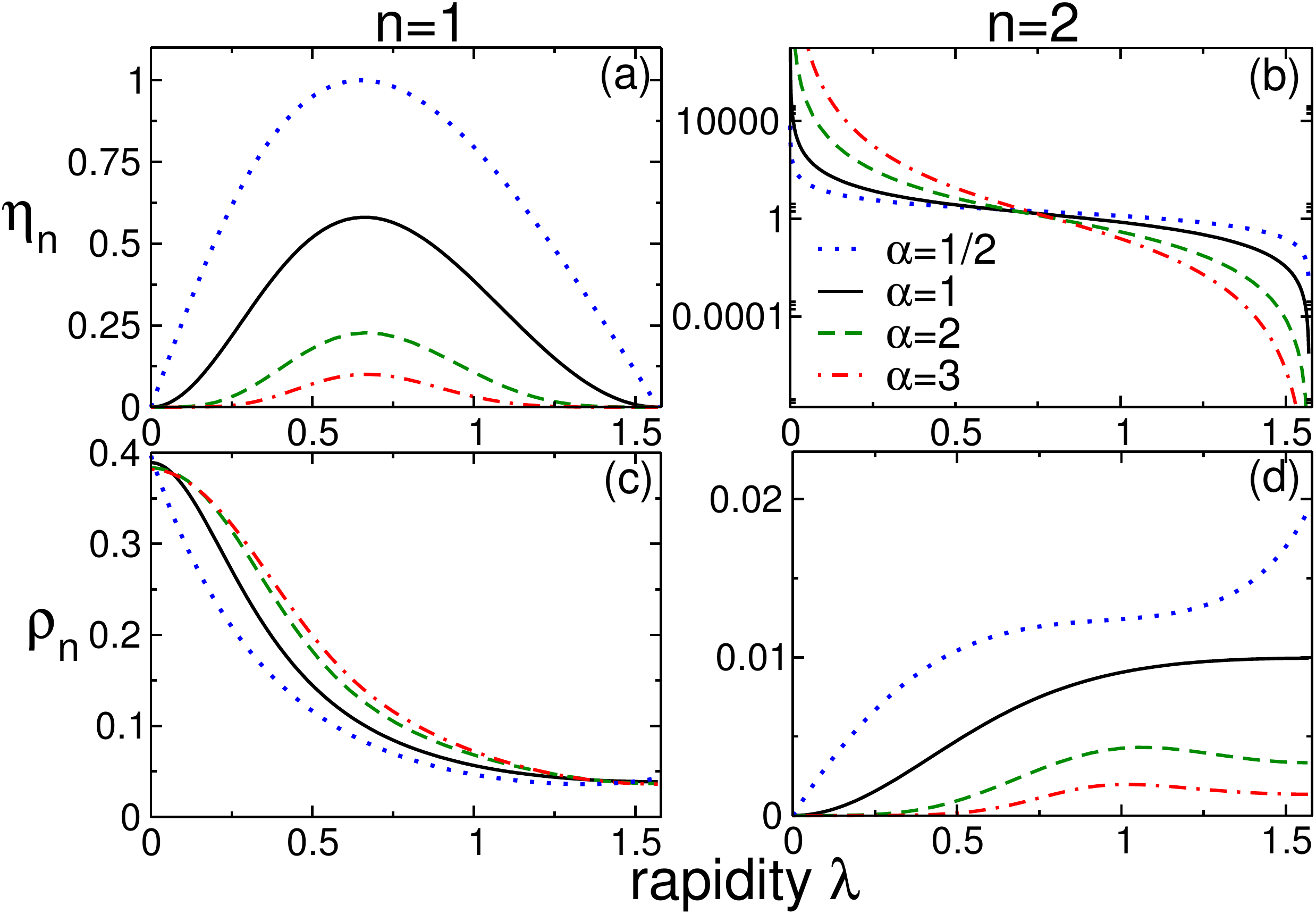}
\end{center}
\caption{ Thermodynamic Bethe Ansatz approach for the diagonal R\'enyi 
 entropies: saddle point rapidity densities $\rho_n,\eta_n$. 
 Because of parity invariance only the interval $[0,\pi/2]$ is plotted. 
 Panels (a) and (b) show $\eta_n$ as a function of the rapidity $\lambda$, for $n=1$ and $n=2$, respectively. 
 Different lines correspond to different R\'enyi  index $\alpha$. 
 The data are numerical solutions of~\eref{tba-eq} and~\eref{qa-tba} for $\Delta=2$. 
 Panels (c) and (d): The same as in (a) and (b) but for the root density $\rho_n$. 
}
\label{dens}
\end{figure}
%

Some numerical results illustrating the qualitative features of the saddle point densities, obtained by solving the coupled
equations~\eref{qa-tba} and~\eref{tba-eq}, are reported in Figure~\ref{dens}. 
The figure shows $\rho_n$ and $\eta_n$  for $n=1,2$, and R\'enyi index $\alpha=1/2,1,2,3$ (different lines in the Figure). 
The data are for $\Delta=2$. For $\alpha=1$ we report the analytically known solution \cite{wdbf-14}.
For any $n$ the densities are invariant under $\lambda\to-\lambda$, reflecting parity invariance.
The qualitative features of  the root densities are very similar with varying $\alpha$. 
For  $\lambda\to 0$,  $\eta_n(\lambda)$ with even $n$ always diverges, as shown for $n=2$.
For $\alpha<1$ the root densities are non-analytic at $\lambda=0$. 

We finally report the form of the R\'enyi diagonal entropies~\eref{d-renyi-main} written explicitly in terms of the root distributions as
\begin{equation}
\label{r-theo-1}
S_d^{(\alpha)}=\frac{L}{1-\alpha}\sum_n\int_0^{\pi/2}d\lambda(
-\alpha\epsilon_n+s_{n}), 
\end{equation}
where
\bea
\label{r-theo-supp}
\epsilon_n&\equiv&\rho_n(g_n+4n\ln2),\\
s_n&\equiv&\rho_n^{(t)}\ln\rho_n^{(t)}-\rho_n\ln\rho_n-\rho_n^{(h)}\ln\rho_n^{(h)}.   
\eea
Here $\epsilon_n(\lambda)$ and $s_{n}(\lambda)$ are the contributions of the bound 
states with rapidity $\lambda$ to the Yang-Yang entropy~\eref{y-y} and to the 
driving ${\cal E}$~\eref{driving}, respectively. In~\eref{r-theo-supp}, $g_n$ is 
as defined in~\eref{gn}, while $\rho_n,\rho_n^{(h)}$ are the saddle point densities 
obtained by solving~\eref{qa-tba} and~\eref{tba-eq}.


\section{Some analytic results}
\label{sec:an}

There are few limiting cases in which the R\'enyi entropies can be calculated analytically or can be analytically related to some 
known results. 
It is very instructive to study in details these limits because they will serve as reference points to check the correctness and accuracy of 
the numerical solutions for the general case and also to give important physical insights about the physics of the stationary R\'enyi entropies. 
Explicitly we consider the limit of large $\Delta$ for arbitrary $\alpha$ and the limits for $\alpha=0,1,\infty$ for arbitrary $\Delta$.

\subsection{Large $\Delta$ expansion}
\label{sec:large-D}

The expansion for large $\Delta$ in the case of a quench from the N\'eel state is non-generic because in the limit $\Delta\to\infty$
the N\'eel state becomes the ground-state of the model and there is no quench. 
Consequently, increasing $\Delta$ all observables approach the ground-state values and in particular the entropies become all zero. 

The expansion for large $\Delta$ is conveniently parametrised in powers of
\be
 z\equiv e^{-\eta}, \qquad {\rm with} \qquad \eta \equiv\textrm{arcosh}(\Delta).
\ee 
As done in Ref.~\cite{wdbf-14} for $\alpha=1$, we use the ansatz for $\eta_n$
\begin{equation}
\label{s-ans}
\eta_n(\lambda)= z^{\beta_n}\eta_n^{(0)}(\lambda)\exp\big[\Phi_n(z,\lambda)\big], 
\end{equation}
where $\beta_n$, $\eta_n^{(0)}$ and $\Phi_n(z,\lambda)$ have to be determined and they all depend  on $\alpha$. 
Plugging this ansatz in~\eref{qa-tba}, and using the small $z$ expansion of the 
driving term $g_n(\lambda)$ and of $a_{nm}$ (see~\eref{gn} and~\eref{anm}, respectively), 
the leading order in $z$ fixes the exponent $\beta_n$ in~\eref{s-ans} as 
\begin{equation}
\label{beta-n}
\beta_n=\left\{
\begin{array}{cc}
0          & n\,\textrm{even},\\
2\alpha    & n\,\textrm{odd}.
\end{array}
\right.
\end{equation}
Moreover, one finds
\begin{equation}
\label{eta-0}
\eta_n^{(0)}(\lambda)=
\left\{\begin{array}{cr}
|\tan(\lambda)|^{-2\alpha}     &   n\,\textrm{even},\\\\
c_n|\sin(2\lambda)|^{2\alpha}  &   n\,\textrm{odd},
\end{array}
\right.
\end{equation}
where the constants $c_n$ are defined as 
\begin{equation}
\label{cn}
c_n\equiv
4^{\alpha}\exp\Big[\frac{1}{\pi(1+\delta_{n,1})}\int_{-\frac{\pi}{2}}^{\frac{\pi}{2}}
\ln(1+|\tan(\lambda)|^{-2\alpha})\Big] .
\end{equation}
Furthermore $\Phi_n(0,\lambda)=0$ so that the small $z$ behaviour is entirely encoded in $\beta_n$ and $\eta^{(0)}_n$.
These results are valid for arbitrary values of $\alpha$.
Eq.~\eref{eta-0} in particular implies that $\eta_n(\lambda)$ diverges in the limit   $\lambda\to 0$ for even $n$, whereas it is vanishing for odd $n$. 
We numerically observed in the previous section that this feature is generic for any finite $\Delta$. 
The behaviour at the origin is determined by the R\'enyi index $\alpha$. 

It is straightforward to show that at this leading order in $z$, the R\'enyi entropies are vanishing.  
Indeed using Eq.~\eref{dec-rho}, we have $\rho_1(\lambda)=1/(2\pi)$, $\rho_{n>1}=\rho^{(h)}_n=0$, which leads to 
${\cal E}(\pmb \rho^*)=S_{YY}(\pmb \rho^*)=0$.
This result reflects the fact that for $\Delta\to\infty$ the N\'eel state is the ground-state of the XXZ Hamiltonian with zero entropy. 
In order to get a non-zero result, we should perform the expansion up to the first non-zero order. 
This is easily done for fixed $\alpha$ (as we will soon do for $\alpha=2$), but it is more cumbersome to analyse generically having 
$\alpha$ as an arbitrary real parameter. 
It is anyhow easy to understand the leading term in $z$ (and hence in $\Delta$).
In fact, for arbitrary $\alpha$, Eq.~\eref{dec-rho} joined with \eref{s-ans} provides
\be\fl\label{rho-lD}
\rho_1(\lambda)=\frac{1}{2\pi}(1+4z\cos(2\lambda)+O(z^2))\,,\qquad \rho_{n>1}= O(z^{2\alpha}),\qquad \rho_{n}^{(h)}=O(z^{2\alpha})\,.
\ee
where many of the terms $O(z^{2\alpha})$ are indeed $o(z^{2\alpha})$. Notice that at this order there is no $\alpha$ dependence. 
Given that up to $O(z)$, $\rho_n=\rho_n^{(t)}$, the corresponding Yang-Yang entropy is vanishing. 
Thus the R\'enyi entropies can potentially get a contribution only from the driving term ${\cal E}$.
However, plugging $\rho_1(\lambda)$ above in \eref{r-theo-supp}, we get ${\cal E}=0$ and so the R\'enyi 
entropies are still all vanishing.  
For $\alpha>1$, the second order (which is $\alpha$ dependent) gives generically a non-zero result.
Again, given that  $\alpha>1$, we can ignore the contributions from $\rho_{n>1}$ and from $\rho_{n}^{(h)}$ which are $O(z^{2\alpha})$.
This again implies that there is no contribution from the Yang-Yang entropy, but that only the second order of $\rho_1(\lambda)$
provides a non-zero contribution from the driving term ${\cal E}$ which is $O(z^2)$.
Thus we generically have that for $\alpha>1$, the R\'enyi entropies are always $O(z^2)$ as we explicitly show for $\alpha=2$ in the following.
For $\alpha=1$, the result has been  worked out in \cite{wdbf-14} and we know $S_{YY}= 4{\cal E}=O(z^2)$,
which is compatible with what derived above. 
Conversely for $\alpha<1$, the terms $z^{2\alpha}$, present in many root and hole densities, matter and the analysis becomes 
more cumbersome. It turns out that the contributions from the driving term and from the Yang-Yang entropy are of the same order, 
as confirmed also by numerical solutions.

Calculating the expansion of $\Phi_n(z,\lambda)$ in~\eref{s-ans} as power series in $z$ is easily done for integer 
$\alpha$ (while it is slightly more cumbersome for real $\alpha)$.
Specifically, at a fixed order in the expansion of the TBA system~\eref{qa-tba} (or~\eref{dec}) one obtains a finite hierarchy 
of equations involving only a finite number of the functions $\Phi_n(z,\lambda)$. 
For simplicity in what follows we exhibit explicit formulas only for $\alpha=2$.
Up to the fourth order in $z$, $\Phi_n$ turn out to be  
\bea
\label{phi}\fl
 && \Phi_1= \ 4z\cos(2\lambda)+4z^2(-2+\sqrt{2})(1-\cos(4\lambda))+z^3\Big(\frac{4}{3}\cos(6\lambda)-12\cos(2\lambda)\Big)
 \\\nonumber \fl && \qquad
-z^4(24(2\sqrt{2} -3)\cos(4\lambda)-4(3\sqrt{2}-4)(3+\cos(8\lambda)))+{\mathcal O}(z^5),\\
 \fl && \Phi_2=   -16z^2\cos(2\lambda)+4z^4(\cos(3\lambda)\sec(\lambda)
+\sin(3\lambda)\csc(\lambda))+{\mathcal O}(z^5),\\
\fl && \Phi_3= \  8z\cos(2\lambda)+8z^2(\sqrt{2}-2)(1-\cos(4\lambda))-z^3\Big(24\cos(2\lambda)-\frac{8}{3}\cos(6\lambda)\Big)
\\\nonumber \fl & & \qquad
-z^4 (8(12\sqrt{2}-17)\cos(4\lambda)-8(3\sqrt{2}-4)(3+\cos(8\lambda)))+{\mathcal O}(z^5),\\
\label{phi1}
\fl && \Phi_{2k}=\Phi_2,\\
\label{phi2}
\fl && \Phi_{2k+1}=\Phi_3. 
\eea
The densities $\rho_n$ are obtained by plugging the expansion for $\eta_n(\lambda)$ into~\eref{tba-eq} (equivalently in~\eref{dec-rho}).
After a straightforward but tedious calculation, we get 
\bea
\label{rho1}
&\rho_1=\frac{1}{2\pi}(1+4z\cos(2\lambda)+4z^2\cos(4\lambda)
-8z^3\sin(2\lambda)\sin(4\lambda)+ \\\nonumber&\qquad
+\frac{z^4}{64}(c_1(32\cos(4\lambda)-21)
-8(c_1-32) \cos(8\lambda)))+{\mathcal O}(z^5),\\
&\rho_2=\frac{3c_1z^4}{32\pi(1+\cot^4(\lambda))}+{\mathcal O}(z^5),\\
&\rho_3=\frac{3c_1z^4}{128\pi}+{\mathcal O}(z^5),\\
&\rho_4=o(z^4),
\eea
where $c_1$ is defined in~\eref{cn} and for $\alpha=2$ it reads $c_1=16 (2+\sqrt{2})$. 
In contrast with $\eta_n$  (cf.~\eref{phi}-\eref{phi2}), the leading order of $\rho_n$ with 
larger $n$ corresponds to higher orders in $z$. In particular, note that $\rho_1={\mathcal O}(1)$, whereas $\rho_2={\mathcal O}(z^4)$,
in agreement with the general analysis exposed above for arbitrary $\alpha$.
%
The hole densities $\rho^{(h)}_n(\lambda)=\rho_n(\lambda) \eta_n(\lambda)$ and the total ones 
$\rho^{(t)}_n=\rho_n(\lambda)+\rho^{(h)}_n(\lambda) $ are straightforwardly obtained.

From these densities we can finally calculate the leading order in $z$ of the Yang-Yang 
entropy associated with the macrostate identified by $\rho_n$ and $\eta_n$ given by 
\bea\fl 
\label{s2}
\frac{S_{YY}}{L}&=&
z^4\frac{c_1}{2\pi}\Big(-4 \ln z \int_{-\frac{\pi}{2}}^{\frac{\pi}{2}} d\lambda \sin^4(2\lambda) 
+ \int_{-\frac{\pi}{2}}^{\frac{\pi}{2}} d\lambda \sin^4(2\lambda) (1-\ln (c_1 \sin^4 (2\lambda)))
\\  \nonumber
\fl &&\qquad + \frac3{16}\int_{-\frac{\pi}{2}}^{\frac{\pi}{2}} d\lambda  \frac{\ln (\cot ^4(\lambda )+1)+\cot ^4(\lambda ) \ln (\tan^4(\lambda )+1)}{\cot^4(\lambda) +1} \Big) +o(z^4)\\
 \fl \nonumber &=&z^4 c_1\Big(-\frac34 \ln z -\frac14-\frac{3\pi}{32\sqrt2}
  \Big)+o(z^4),
\eea
which is compatible with the general behaviour $z^{2\alpha}$.
In Eq. \eref{s2} some miraculous cancellations between the terms coming from $\rho_1$ and $\rho_2$ happen, 
signalling that there could be some hidden structure.    
The leading contribution of the driving term ${\cal E}$ is 
\bea
\label{d2}
\frac{{\mathcal E}}{L}&=&
\frac{1}{\pi}\int_{0}^{\frac{\pi}{2}}d\lambda\Big\{\frac{1}{4}\ln(4\sin^2(2\lambda))
+z\cos(2\lambda)(1+\ln4+\ln(\sin^2(2\lambda)))+\\ && \nonumber \qquad+
z^2\Big[2+\cos(4\lambda)\big(2+\ln4+ \ln(\sin^2(2\lambda))\big)\Big]\Big\}\\ 
&=&\frac{z^2}{\pi}\int_{0}^{\frac{\pi}{2}}d\lambda \Big[2+2 \cos(4\lambda) \ln \sin(2\lambda)\Big]=\frac{z^2}2\,,
\eea
where we used that the first two orders in~\eref{d2} and some parts in the third order vanish.
The leading orders of $S_{YY}$ and ${\cal E}$ are then given by different powers of $z$ being $z^4 \ln z$ and $z^2$ respectively. 
Consequently, for large $\Delta$,  $S_{YY}$ is subleading compared with ${\cal E}$, implying that the R\'enyi diagonal entropy for $\alpha=2$ 
is dominated by the driving term in the limit $z\to 0$, as we have generically shown to be the case for $\alpha>1$. 

A final observation is now in order.  The function which is integrated in Eq. (\ref{s2}) to get $S_{YY}$ is positive for any $\lambda$.
Conversely, the function integrated for ${\cal E}$ in Eq. \eref{d2} is negative for some values of $\lambda$. 
Since the latter dominate the sum, we have that the integrated function cannot be considered as the contribution of the Bethe mode
with momentum $\lambda$ to the entropy, because this must be a positive function.

\subsection{The max entropy, i.e. the limit $\alpha\to0$ }
\label{sec:tba-renyi-1}

It is instructive to explicitly consider the limit $\alpha\to 0$ of Eq.~\eref{r-theo-1} defining the max entropy, 
which counts the number of eigenstates of the $XXZ$ chain with non-zero N\'eel overlap. 
In this case, the TBA equations \eref{qa-tba} become those of the thermal ensemble at infinite temperature (i.e. $\beta=0$, 
see e.g. \cite{taka-book} for comparison) and so the diagonal entropy is 
\be
S_d^{(0)}=L \frac{\ln 2}2 \,,
\ee
i.e. half the Yang-Yang entropy of the thermal ensemble at infinite temperature. 
The factor $1/2$ in the exponent reflects that only parity-invariant eigenstates can have non-zero N\'eel overlap. 
This is in agreement with the well known fact that the total number of eigenstates 
with non-zero N\'eel overlap, for large $L$ scales like $\propto 2^{L/2}$ \cite{ac-16qa}  
(the total number of eigenstates with non-zero  overlap with the N\'eel state has been obtained analytically at $\Delta=1$ \cite{ac-16qa}
and at $\Delta=0$ \cite{msca-16}).

\subsection{The von Neumann entropy, i.e. the limit $\alpha\to1$ }

In \cite{ac-17} it has been shown in full generality that the diagonal entropy at $\alpha=1$ is half of the  Yang-Yang entropy.
It is important to recover this result from the limit $\alpha\to 1$ of ~\eref{r-theo-1} and \eref{qa-tba} to show the self-consistency of 
our approach. However, we relegate this test to \ref{app1} because it does not provide any new physical insight.

\subsection{The min entropy, i.e. the limit $\alpha\to\infty$}
\label{sec:min-ent}

We now analyse the min entropy which is obtained by taking the limit $\alpha\to\infty$ of the R\'enyi entropies~\eref{d-renyi}. 
The same limit for the entanglement R\'enyi entropies defines the single copy entanglement \cite{orus-2006}. 
Similar to finite $\alpha$, the min entropy exhibits volume-law behaviour. 
Its density is given in terms of a thermodynamic macrostate that here we determine analytically. 
As clear from the definition of the diagonal min entropy, at a microscopic level this state is the eigenstate with the largest N\'eel overlap. 
Interestingly, we observe two different regimes. For large $\Delta$ the macrostate coincides with the ground state of the $XXZ$ chain. 
As it is well known, this has zero Yang-Yang entropy and it contains only one-strings (i.e. $n$-strings with $n>1$ are not present). 
Oppositely, at low $\Delta$ the macrostate is an excited state. We find that this has still zero Yang-Yang entropy but it 
contains non-trivial bound states. The transition between the two behaviours happens at a special value of $\Delta$ that we determine. 

In order to understand the presence of these two regimes, it is instructive to check what happens if one takes the limit $\alpha\to\infty$
of the large $\Delta$ expansion of Sec. \ref{sec:large-D}. 
Because of the term $(\sin2\lambda)^{2\alpha}$ in \eref{eta-0} present in $\eta_n$ for $n$ odd, one could naively expect
$\eta_n(\lambda)\to0$ for $\alpha\to \infty$ (unless $\lambda=\pi/4$, but this is a measure zero point). 
But this is not the case, because one should also check that the constants $c_n$ in \eref{eta-0} stay finite as $\alpha\to\infty$.
This is not the case and indeed $c_n$ can diverge because 
\begin{equation}\fl
\int_{-\frac{\pi}{2}}^{\frac{\pi}{2}}\ln(1+|\tan(\lambda)|^{-2\alpha})d\lambda= 4\alpha G+o(\alpha),
\quad \Rightarrow\qquad
c_{n>1}\simeq (2 z e^{2G/\pi})^{2\alpha},
\end{equation} 
where $G=0.915\dots$ is the Catalan constant.  
This implies that it is possible to have 
 \begin{equation}
\label{cond}
\lim_{\alpha\to\infty}\eta_n(\lambda)\to 0,\quad\forall \,n\,\textrm{odd}. 
\end{equation}
for any $\lambda$ only if (assuming that all $\Phi_n(z,\lambda)$ in~\eref{s-ans} do not diverge as $\alpha\to\infty$)
\begin{equation}
\label{crit}
\ln(z)\le -\ln(2)-2\frac{G}{\pi}. 
\end{equation}
The condition~\eref{crit} is clearly satisfied for small enough $z$ (i.e. large enough $\Delta$), but it also 
suggests that there is a ``critical'' anisotropy $\Delta^*\approx 1.93$ above which~\eref{cond} holds for any $\lambda$. 
We are going to  show that this result is qualitatively correct, although, due to the large 
$\Delta$ approximation, the value of $\Delta^*$  from \eref{crit} is not accurate.

\subsubsection{The min entropy for $\Delta> \Delta^*$ and the determination of $\Delta^*$.}
\label{largeD}

In order to determine the analytic behaviour of the min entropy for $\Delta>\Delta^*$ and to self-consistently determine the 
value of $\Delta^*$, the trick is to impose that~\eref{cond} is satisfied and check that this is indeed possible and that allows us to solve 
the TBA equations~\eref{dec}. We stress that it would have been very unlikely to get the idea of imposing the condition \eref{cond} without 
having first performed the large $\Delta$ expansion. 

The equations~\eref{cond} are recursive and couple each $\eta_n$ with $\eta_{n\pm1}$. Clearly if all the odd $\eta_n$ are vanishing 
according to the condition \eref{cond}, these equations decouple and in the limit $\alpha\to\infty$, they simplify to
\begin{equation}
\label{rrr}
\ln\eta_n=\left\{
\begin{array}{cc}
\alpha d_2 & n\,\textrm{even},\\
\alpha d_1 +s\star\ln(1+e^{\alpha d_2})& n=1,\\
\alpha d_1+2s\star\ln(1+e^{\alpha d_2}) & n\,\textrm{odd},
\end{array}
\right.
\end{equation}
where the functions $d_n$ are defined in \eref{dn}.
Eq. (\ref{rrr}) for $n$ odd must be intended as a self-consistent equation with the condition \eref{cond}, i.e. 
that the rhs should go to $-\infty$ as $\alpha\to\infty$. 
In this respect, the fact that $d_1(\lambda)<0\,\forall\lambda$ for any value of $\Delta$ pushes this rhs in the right direction 
and we are only left with the analysis of the convolution integral. in~\eref{rrr}.
The latter can be simplified as follows. Since $d_2(\lambda)>0$ for $\lambda\in [-\pi/4,\pi/4]$ 
(which is the only relevant integration region in the limit $\alpha\to\infty$) we can rewrite, as $\alpha\to\infty$, the convolution as 
\begin{equation}
s\star\ln(1+e^{\alpha d_2})=\alpha s\star d_2(\lambda)\theta(|\lambda|-\pi/4). 
\end{equation}
At this point, in order to be consistent with~\eref{cond}, one requires  
\begin{equation}
\label{cond1}
d_1+2 s\star d_2<0, \qquad {\rm and} \qquad d_1+ s\star d_2<0.
\end{equation}
These inequalities can be analysed using
\begin{equation}
d_n=\sum\limits_{k\in\mathbb{Z}}e^{-2i k\lambda}\frac{\tanh(k\eta)}{k}[(-1)^n-(-1)^k], 
\end{equation}
which leads to
\bea\fl
\label{cv}
s\star d_2=\frac{1}{4}\sum\limits_{k\in\mathbb{Z}}
\Bigg[ e^{-2ik\lambda}\frac{\tanh(k\eta)}{k\cosh(k\eta)}
[1-(-1)^k]+\\
+\sum\limits_{k\ne k'}e^{-2ik\lambda}\frac{\sin((k-k')\pi/2)}{2\pi(k-k')}
\frac{\tanh(k'\eta)}{k'\cosh(k\eta)}(1-(-1)^{k'}) \Bigg]. \nonumber
\eea
By using~\eref{cv} and the expression for $d_1(\lambda)$ one can work out numerically 
that~\eref{cond1} holds for $\Delta>\Delta^*=1.76692...$, which is slightly lower than the result~\eref{crit}
from the large $\Delta$ expansion. 

%
\begin{figure}[t]
\begin{center}
\includegraphics*[width=0.87\linewidth]{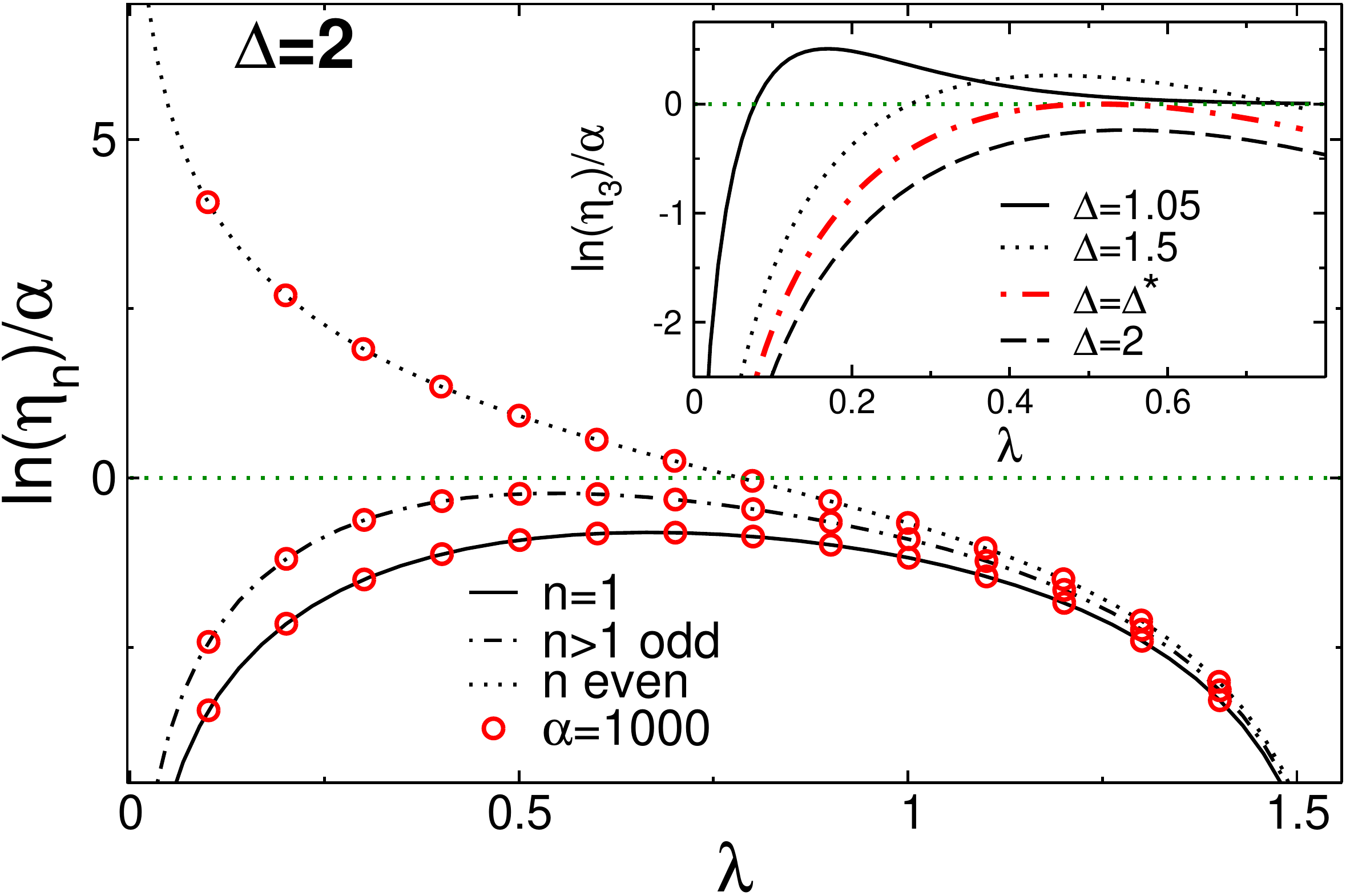}
\end{center}
\caption{Diagonal R\'enyi entropies in the limit 
 $\alpha\to\infty$: saddle point densities $\eta_n$ at 
 $\Delta=2$. Main figure: $\ln(\eta_n)/\alpha$ plotted as a function 
 of the rapidity $\lambda$. The lines are the analytical results~\eref{rrr}. 
 The circles are obtained by solving numerically the TBA equations for 
 $\alpha=10^3$. For odd $n$ one has $\eta_n<0$ for any $n$. Note also that 
 $\eta_n>\eta_1$ for $n>1$. For  even $n$, $\eta_n$ changes sign at 
 $\lambda=\pi/4$, and it diverges at small $\lambda$. Inset: The 
 special point $\Delta^*\approx 1.7669$ at which $\ln(\eta_n)/\alpha$ 
 (dash-dotted line) touches the real axis.  
}
\label{check_D2}
\end{figure}
%

The results~\eref{rrr} for $\eta_n$ are shown in Figure~\ref{check_D2}. 
Since $\eta_n$ are even functions of $\lambda$, we restrict ourselves to the region $\lambda>0$.  
The Figure reports $\ln(\eta_n)/\alpha$ as functions of $\lambda$ for $\Delta=2$ (plots for other values of $\Delta>\Delta^*$ are equivalent). 
The continuous curves correspond to the analytic solution~\eref{rrr} while the circles are obtained by solving numerically the 
TBA equations for $\alpha=10^3$: the two are in perfect agreement. 
In particular, one has that $\ln(\eta_n)/\alpha<0$ for $n$ odd, in agreement with~\eref{cond}, 
while for even $n$, $\ln(\eta_n)$ is positive for $|\lambda|<\pi/4$, and it diverges at $\lambda\to 0$. 
Note also that for any $\lambda$ one has $\eta_3>\eta_1$ (we recall that all $\eta_n$ with $n$ odd and $n\neq1$ are equal). 
The inset in the  Figure shows $\ln(\eta_3)/\alpha$ for various $\Delta$ both larger and smaller than  $\Delta^*\approx 1.7669$ (cf.~\eref{cond1}). 
For $\Delta<\Delta^*$ there is an extended region where $\ln(\eta_3)>0$, implying that the condition~\eref{cond} is violated and 
the solution~\eref{rrr} is not valid. For $\Delta=\Delta^*$, $\eta_3(\lambda)$ is tangent to the horizontal axis. 
The results for $\eta_n$ in the region with $\Delta<\Delta^*$ are discussed in~\ref{smallD}. 


We are now ready to derive analytically $\rho_n$ for $\Delta\ge\Delta^*$. 
Again, it is instructive to look at what happens in the large $\Delta$ limit (see~\eref{rho1}).
As discussed above, we have $\rho_1={\mathcal O}(1)$ and $\rho_{n>1}\sim O(z^{2\alpha})$. 
Thus in the limit  $\alpha\to\infty$ we have
\begin{equation}
\label{rho-n}
\rho_n\to 0,\quad n\ge 2. 
\end{equation}
On the other hand, from~\eref{tba-eq}, $\rho_1$ is determined by solving 
the integral equation 
\begin{equation}
\label{a-inf}
\rho_1=a_1-a_{11}\star\rho_1. 
\end{equation}
To derive~\eref{a-inf} we used that $\rho_1^{(h)}\to 0$ because $\eta_1\to 0$ (cf.~\eref{rrr}) 
and $\rho_1$ is assumed to be regular. 
Eq.~\eref{a-inf} is the same integral equation that identifies the ground state root density of the $XXZ$ chain~\cite{taka-book} which 
is solved by Fourier transform providing 
\begin{equation}
\label{rho-1}
\rho_1=s(\lambda)=\frac{1}{2\pi}\sum \limits_{k\in\mathbb{Z}}\frac{e^{2 i k\lambda}}{\cosh(k\eta)}. 
\end{equation}
The energy of the state reads 
\begin{equation}
\frac{E}{L}=-\sinh(\eta)\sum\limits_{k\in\mathbb{Z}}\frac{1}{e^{2|k|\eta}+1}. 
\end{equation}

Eq.~\eref{rho-1} and~\eref{rho-n} can be also found without relying on the small $z$ expansion.  
Indeed, plugging the condition \eref{cond} in the recursive equations~\eref{dec-rho} 
we have for the even densities
%
$\rho_{2k}(\lambda)(1+\eta_{2k}(\lambda))\to 0$, i.e.  $\rho_{2k}(\lambda)\to0\forall k$. 
%
Plugging this result for even $n$ in~\eref{dec-rho}, we obtain also the odd densities as $\rho_{2k+1}(\lambda)\to 0$ for $k>0$, and 
\begin{equation}
\rho_1(\lambda)=s(\lambda), 
\end{equation}
which is the same as \eref{rho-1} and correspondst to the ground-state of the XXZ spin-chain.

%
\begin{figure}[t]
\begin{center}
\includegraphics*[width=0.87\linewidth]{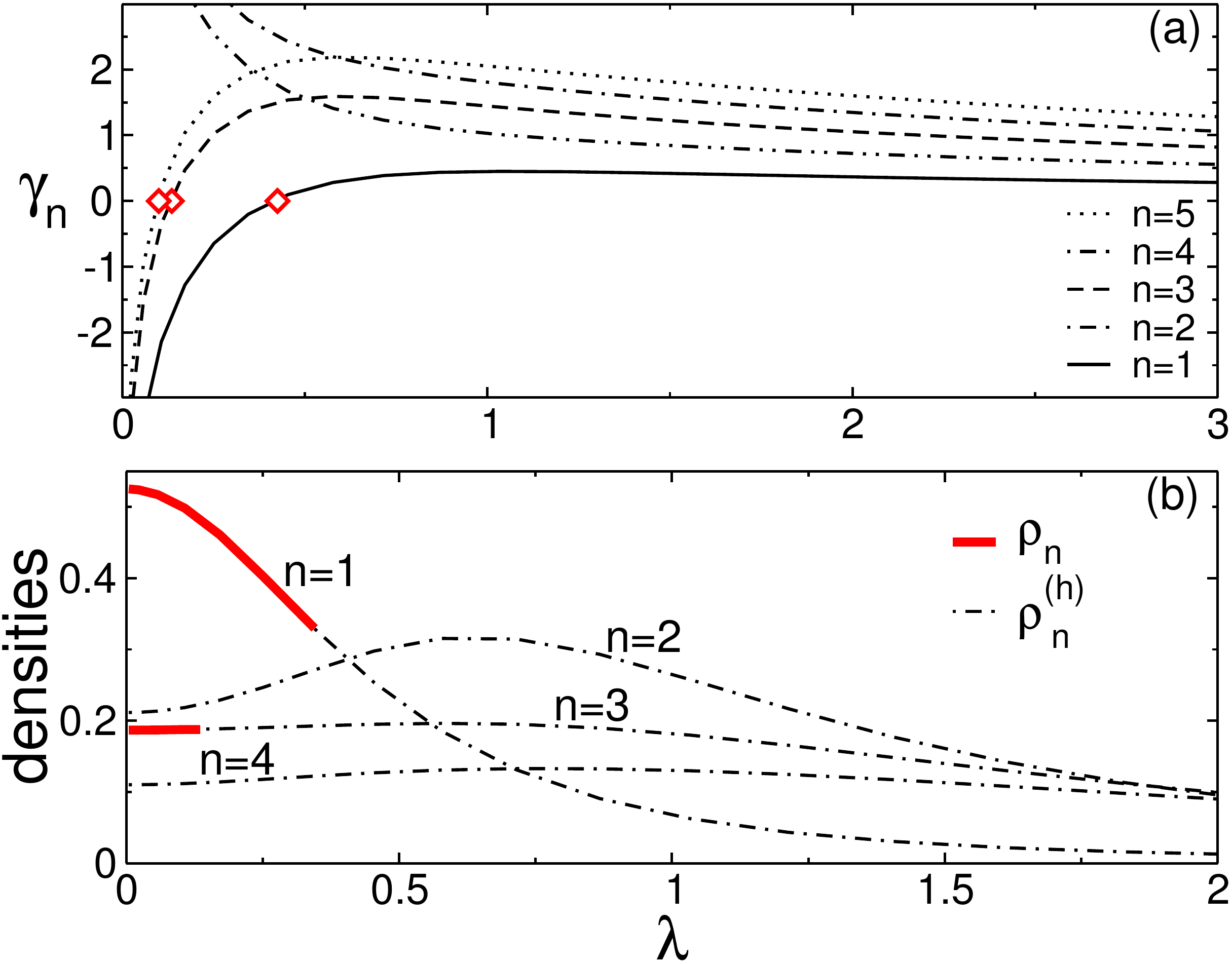}
\end{center}
\caption{Diagonal R\'enyi entropies in the limit $\alpha\to\infty$ in the $XXX$ chain. 
 Panel (a): Saddle point $\gamma_n\equiv\ln(\eta_n)/
 \alpha$ as a function of rapidity $\lambda$. For even $n$, $\gamma_n$ 
 are positive and diverge at small $\lambda$, while they vanish 
 for $\lambda\to\infty$. For odd $n$, $\gamma_n\to-\infty$ at small 
 $\lambda$. The diamonds mark the points where $\gamma_n$ change sign. 
 Panel (b): Particle densities $\rho_n$ (full lines) and hole densities $\rho_n^{(h)}$ (dash-dotted lines). 
 Only the non-zero values of $\rho_n$ and $\rho_n^{(h)}$ 
 are shown. Note that both densities are not continuous functions. $\rho_n$ 
 is non-zero only for odd $n$. For generic $n$, $\rho_n$ 
 ($\rho_n^{(h)}$) is non-zero only if $\gamma_n$ is positive  
 (negative) (see Figure~\ref{check_D2}). 
}
\label{eta_odd}
\end{figure}
%

\subsubsection{The min entropy for $\Delta< \Delta^*$.}
\label{smallD}

For $\Delta<\Delta^*$ it is convenient to use the parametrisation 
\begin{equation}
\label{etan}
\eta_n=\exp(\alpha\gamma_n). 
\end{equation}
Here the finite functions $\gamma_n(\lambda)$ have to be determined by solving the 
TBA system~\eref{dec} for $\eta_n$. In the limit $\alpha\to\infty$ the 
equations for $\gamma_n$ are obtained from~\eref{dec} as 
\begin{equation}
\label{alpha-eta}
\gamma_n=d_n+s\star [\gamma^+_{n-1}+\gamma^+_{n+1}],\quad
\gamma^+_n(\lambda)\equiv\left\{
\begin{array}{cc}
\gamma_n(\lambda)         & {\rm if}\;  \gamma_n(\lambda)>0,\\
0   & {\rm if}\; \gamma_n(\lambda)<0.
\end{array}
\right.
\end{equation}
In the right-hand side only the positive part $\gamma^+_n$ of $\gamma_n$ appear, which makes the equations non linear. 
The values of $\lambda$ where $\gamma_n$ change 
sign have also to be determined from~\eref{alpha-eta}. 

The fact that $\eta_n(\lambda)$ either diverges or vanishes in the large $\alpha$ limit (except in special points when $\gamma_n=0$)
implies that $\rho_n(\lambda)$ and $\rho^{(h)}_n(\lambda)$ have complementary domains in which they are non zero
(under the reasonable assumption that they are finite, except in isolated points). 
In particular $\rho_n(\lambda)$ is non-zero only for those $\lambda$ such that $\gamma_n(\lambda)<0$ and viceversa for $\rho^{(h)}_n(\lambda)$.
Because of the complementarity of the domains, the TBA equations~\eref{tba-eq} for $\rho_n(\lambda)$ and $\rho^{(h)}_n(\lambda)$
decouple. 
First the non zero-part of $\rho_n(\lambda)$ is obtained by solving the equations 
\begin{equation}
\label{rhoplus}
\rho_n(\lambda)=a_{n}(\lambda)-\sum\limits_{m=1}^\infty(a_{nm}\star\rho_m)(\lambda),
\end{equation}
where each $\rho_m$ is non zero only where $\gamma_n(\lambda)<0$.
Similarly, $\rho_n^{(h)}$ is non-zero only on the support of $\gamma^+_n$. 
From the solutions $\rho_n$ of ~\eref{rhoplus} $\rho_n^{(h)}$ is obtained as
\begin{equation}
\label{rhoholes}
\rho^{(h)}_n=a_{n}-\sum\limits_{m=1}^\infty(a_{nm}\star\rho_m),
\end{equation}
which seems the same as \eref{rhoplus}, but it is defined in the complementary domain. 

Equations \eref{alpha-eta}, \eref{rhoplus}, and \eref{rhoholes} cannot be handled analytically, but are easily  solved numerically. 
Their numerical solutions for $\gamma_n$ and $\rho_n$ are reported in Figure~\ref{eta_odd} for $\Delta=1$ 
(for which $\lambda$ is defined on the entire real axis).   
Panel (a) shows $\gamma_n$ for $n\le 5$ as obtained by numerically solving~\eref{alpha-eta}. 
Due to the parity symmetry, we only show results for $\lambda\ge0$. 
For even $n$, $\gamma_n$ are positive for any $\lambda$, and they diverge as $\lambda\to 0$ so that 
$\rho_{2k}\to 0$ in the same limit.  
On the other hand, for odd $n$, $\gamma_n\to-\infty$ at small $\lambda$, 
whereas $\gamma_n$ is positive for large enough $\lambda$. Thus $\gamma_n$ for odd $n$ must change sign at least once:
the points where this happens are marked with the diamonds in Figure~\ref{eta_odd} (a). 

Numerical results for $\rho_n$ are reported in Figure~\ref{eta_odd} (b). 
The continuous lines are the particle densities $\rho_n$, while the dash-dotted 
lines are the hole densities $\rho_n^{(h)}$. For even $n$, $\rho_n$ is identically 
zero, as expected, while for odd $n$ particle and hole densities have 
complementary support. 
Thus for any $n$, $\rho_n^{(t)}(\lambda)$ is either equal to $\rho_n^{(h)}(\lambda)$ or to $\rho_n(\lambda)$ and so 
the thermodynamic macrostate has zero  Yang-Yang entropy.

\subsubsection{Results for the min entropy.}
Given that in both regimes $\Delta>\Delta^*$ and $\Delta<\Delta^*$ the Yang-Yang entropy of the macroscopic state is zero, 
the min entropy is just given by the overlap (driving) term \eref{r-theo-supp} as
\be
S_d^{(\infty)}=L\sum_n\int_0^{\pi/2}d\lambda \rho_n(\lambda)(g_n(\lambda)+4n\ln2)  , 
\ee
with $g_n$ defined in \eref{gn}. Thus, as it should, the min entropy identifies a single eigenstate with the largest overlap
that for $\Delta>\Delta^*$ is the ground-state.
The $\Delta$ dependence of the min entropy is reported in Figure \ref{theo} (together with the results for other values of $\alpha$).
At $\Delta=\Delta^*$ there is a transition between two curves which anyhow is very smooth.
The inset of the Figure compares for $\Delta<\Delta^*$ the logarithm of the overlap of the ground-state (i.e. the analytic 
continuation of the curve for $\Delta>\Delta^*$) with the actual min entropy, showing that the difference is sizeable only 
for $\Delta$ very close to $1$.

The energy of the state with the largest overlap is shown in Figure \ref{ener}  (again together with the results for other values of $\alpha$).
For $\Delta\ge\Delta^*$, this is the energy of the ground state of the $XXZ$ chain (shown as dash-dotted line in the Figure). 
The small circles for $\Delta<\Delta^*$ are the results obtained using the saddle point densities in~\ref{smallD} 
and again the transition between the two regimes is very smooth.  
The largest difference is at $\Delta=1$ for which $E/L\approx-0.66$ compared to the ground state energy density $-\ln2\approx-0.693$. 

\section{General results for the R\'enyi entropies} 
\label{sec:de-sum}

%
\begin{figure}[t]
\begin{center}
\includegraphics*[width=0.95\linewidth]{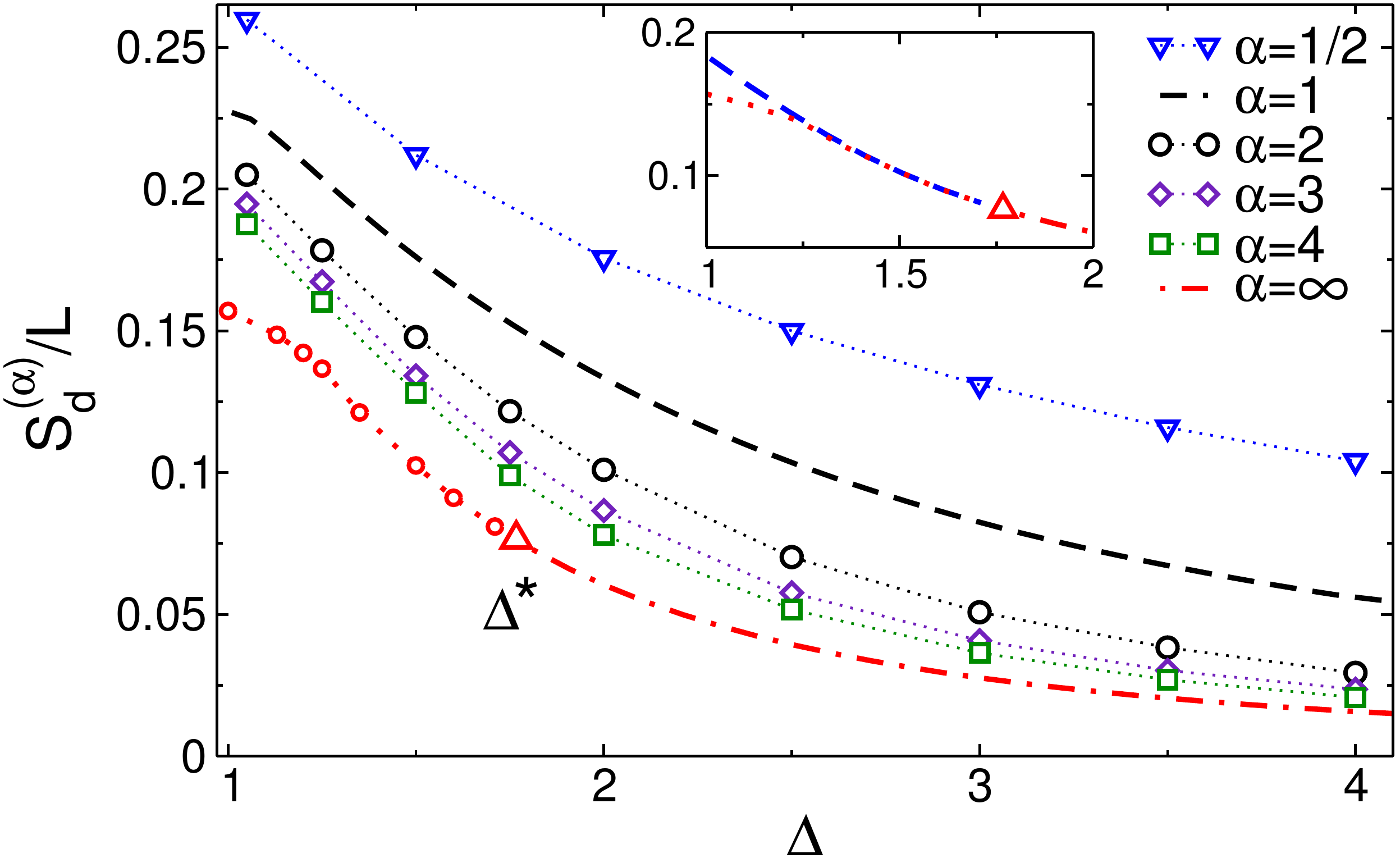}
\end{center}
\caption{Thermodynamic  Bethe Ansatz results for the  Diagonal R\'enyi entropy density $S_{d}^{(\alpha)}/L$ 
 after the quench from the N\'eel state in the $XXZ$ chain as a function of the anisotropy $\Delta$. 
 The dashed line is the von Neumann entropy for $\alpha=1$. Different 
 symbols are used for different $\alpha$. The result for $\alpha\to\infty$ is also shown (the triangle marks the special point $\Delta^*$). 
 Inset:  $S^{(\infty)}_d/L$ for $\Delta<\Delta^*$. The dotted line is the result 
 in the main Figure. The dashed line is obtained using the same saddle 
 point densities as for $\Delta>\Delta^*$. 
}
\label{theo}
\end{figure}
%
In this section we report the results for the diagonal R\'enyi entropies $S_d^{{}_{(\alpha)}}$ which are half of the 
thermodynamic ones. 
Figure~\ref{theo} shows numerical results obtained from Eq.~\eref{r-theo-1} in which we plugged the solution
for the root densities of the TBA equations \eref{qa-tba} (or equivalently \eref{dec}) for the quench from the N\'eel  state. 
The entropy densities $S_d^{(\alpha)}/L$ are plotted as a function of the chain anisotropy $\Delta$. 
%
All the entropies are vanishing in the limit $\Delta\to\infty$, because the N\'eel state is the ground state of the $XXZ$ chain in that limit
and in agreement with the result of the previous section. 
For $\alpha=1$ the data correspond to $S_{YY}/2$ which is obtained by using the analytical results for the thermodynamic macrostate 
in Ref.~\cite{wdbf-14}. 
From the Figure it is clear that $S_d^{{}_{(\alpha)}}\le S_d^{{}_{(\alpha')}}$ for $\alpha'<\alpha$, as expected. 
For $\alpha=\infty$, the dash-dotted line is the result for $\Delta>\Delta^*$ (obtained in section~\ref{largeD}), whereas the small circles are for 
$\Delta\le\Delta^*$ (see~\ref{smallD}). The value of $\Delta^*\approx 1.7669$ is marked by the triangle. 
%

A last consistency check is provided by the general inequality 
\begin{equation}
\label{ineq}
S_d^{(2)}=-\ln\sum_n w^2_n\le -2\ln(\max\limits_n w_n)=2S_d^{(\infty)}, 
\end{equation}
which is satisfied by all our data. 
From the TBA point of view, this is a non trivial check because it is based on the solution of different TBA equations.  
We numerically observe that~\eref{ineq} is saturated for large $\Delta$ because
the sums in~\eref{ineq} are dominated by the largest overlap, which is separated by a gap from the smaller overlaps.

\subsection{Energy of the macrostate} 
\label{sec:ener}

%
\begin{figure}[t]
\begin{center}
\includegraphics*[width=0.91\linewidth]{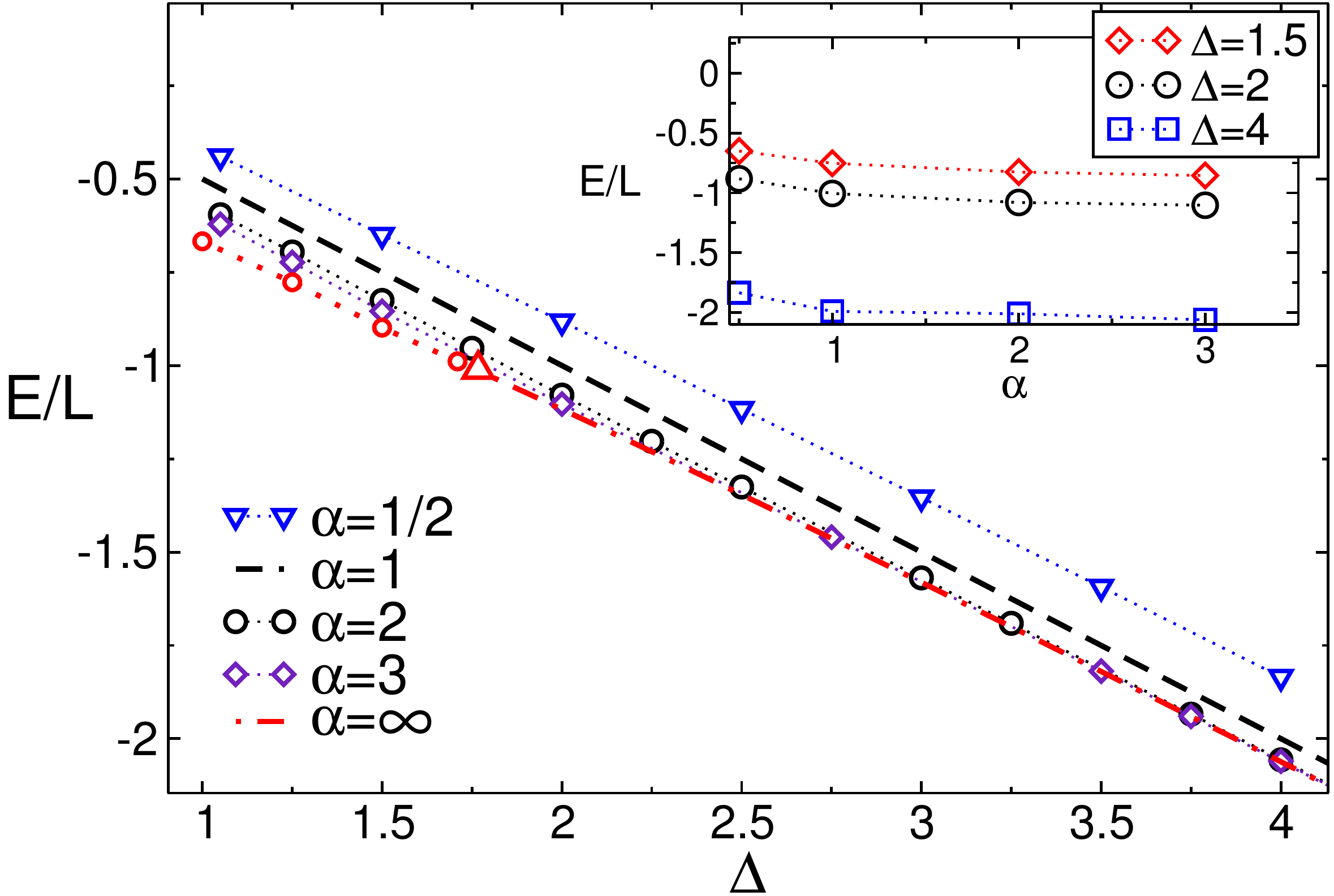}
\end{center}
\caption{ Energy density $E/L$ of the saddle point representative state for the  diagonal R\'enyi entropies plotted versus the 
 chain anisotropy $\Delta$ for several values of the R\'enyi index  $\alpha$. 
 The dashed line is the exact result $E/L=-\Delta/2$ for $\alpha=1$. 
 On the line  $\alpha\to\infty$ the triangle marks the special point $\Delta^*$. 
 The inset shows the $\alpha$ dependence of $E/L$ for three values of $\Delta$.
}
\label{ener}
\end{figure}
%
It is an interesting exercise to investigate the energy density of the macrostate $\rho_n$ because  it 
depends on $\alpha$, implying that it is different from that describing the stationary behaviour of local observables. 
The latter is recovered in the limit  $\alpha=1$. The energy of the macrostate provides information about the 
region in the energy spectrum of the $XXZ$ model that is relevant to describe the R\'enyi diagonal entropies. 
We have already seen that $\alpha=0$ is equivalent to the infinite temperature state, while $\alpha\to\infty$ approaches the 
ground state of the model (at least for $\Delta>\Delta^*$), i.e. zero temperature. 
Hence, by varying $\alpha$ we explore the entire energy window relevant for the XXZ spin-chain.

Our results are shown in Figure~\ref{ener}, reporting the energy density $E/L$ of the macrostate as a function of $\Delta$. 
This is readily obtained plugging the densities $\rho_n$ (cf.~\eref{tba-eq} and~\eref{qa-tba}) in~\eref{c-quant}. 
The symbols are results for different values of $1/2\le\alpha\le\infty$. The dashed line is for $\alpha=1$, which corresponds to 
$E/L=\langle\Psi_0|H|\Psi_0\rangle=-\Delta/2$. 
For all values of $\alpha$, the large $\Delta$ behaviour is straightforwardly calculated plugging \eref{rho-lD} into~\eref{eps} obtaining
\begin{equation}
\label{e-pert}
\frac{E}{L}=
-\frac\Delta2-\frac1{4\Delta}+O(\Delta^{-2}), 
\end{equation}
independently from $\alpha$. Higher orders in $\Delta$ do depend on $\alpha$. 
The explicit $\alpha$ dependence is reported in the inset for fixed value of $\Delta$. 
Form this inset, it is clear that the various curves are very similar and the main difference is the shift in energy.

\subsection{Numerical checks} 
\label{sec:checks}

In this section we provide numerical evidence for the results presented in section~\ref{sec:tba-renyi}. 
We employ two different methods: in subsection~\ref{sec:ed} by using exact (full) diagonalisation techniques we construct explicitly the 
diagonal ensemble~\eref{rho-diag} and the diagonal R\'enyi entropies (cf.~\eref{d-renyi}) for chains of length $L\le 22$; 
in subsection~\ref{sec:ba-num} the diagonal R\'enyi entropies are obtained numerically by exploiting the knowledge of the 
overlaps between the N\'eel state and the Bethe eigenstates, following the approach of Ref.~\cite{ac-16qa}. 

\subsubsection{Exact diagonalisation.}
\label{sec:ed}
%
\begin{figure}[t]
\begin{center}
\includegraphics*[width=0.95\linewidth]{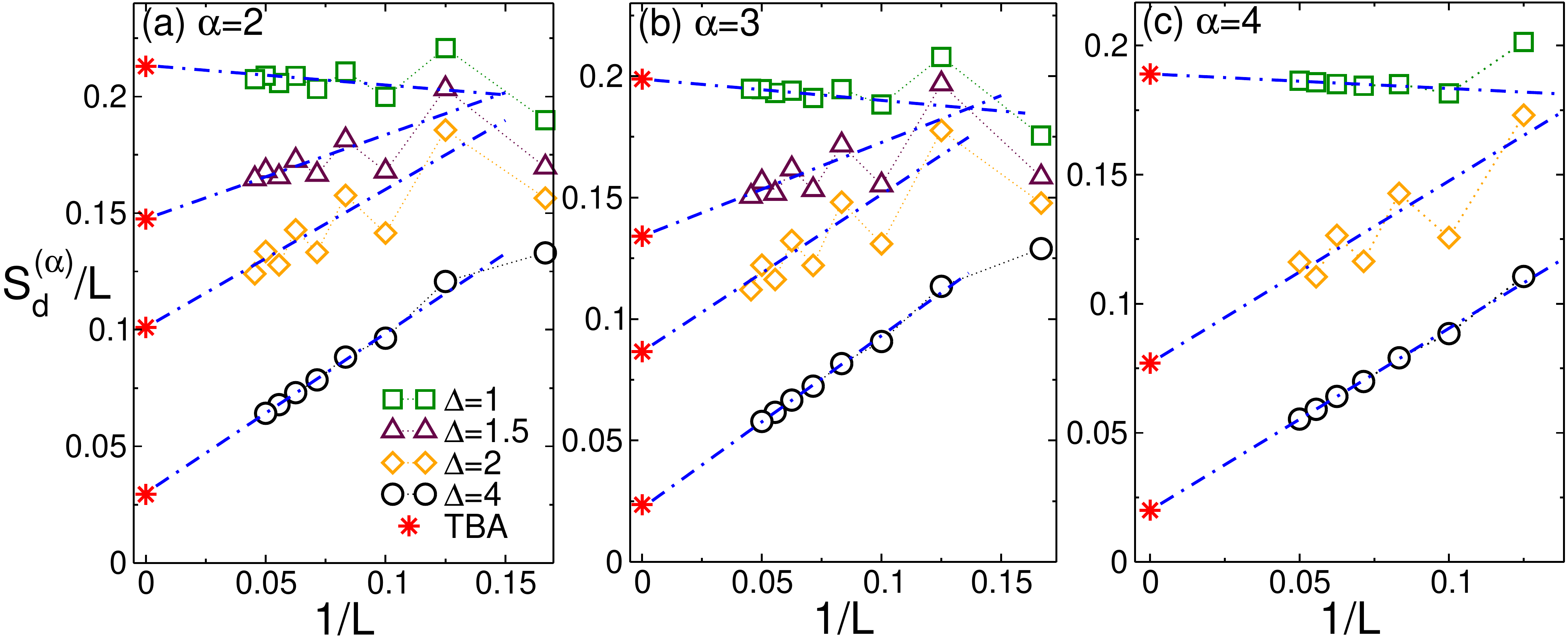}
\end{center}
\caption{ Diagonal R\'enyi entropies after the quench from  the N\'eel state for several values of $\Delta$: $S_{d}^{ (\alpha)}/L$ plotted 
 against the inverse chain length $1/L$. 
 Panels (a), (b) and (c) correspond to $\alpha=2$, $\alpha=3$, and $\alpha=4$ respectively. The stars are the TBA results in the thermodynamic limit. 
 The dash-dotted lines are fits to $S_d^{(\alpha)}/L= s^{(\alpha)}_{\infty}+b_\alpha/L$, with $s_{\infty}^{(\alpha)}$ 
 fixed to the Bethe ansatz result, and $b_\alpha$ fitting parameters. 
}
\label{diag-ed}
\end{figure}
%

The symmetric N\'eel state \eref{in-state} has zero magnetisation, and it 
is invariant under both one-site translations and under parity inversion. Thus, only eigenstates 
in the sector with zero magnetisation, zero momentum, and 
invariant under parity can have non zero N\'eel overlap (see Ref.~\cite{sandvik-2010} 
for the implementation of these symmetries in exact diagonalisation). Here we restrict ourselves 
to this sector of the Hilbert spaces. For $L=22$ this contains 
$N=16159$ eigenstates. This is a small fraction of the 
total number of eigenstates $2^{22}\sim 4\cdot 10^6$ of the $XXZ$ chain, 
although it is still quite large compared with the number of parity-invariant 
eigenstates (i.e. the only ones  with non-zero N\'eel overlap),  
which is $\sim 2^{L/2-1}\sim 500$~\cite{msca-16,ac-16qa}.  
In section \ref{sec:ba-num} by exploiting this property within the formalism of Bethe 
ansatz, we will construct the diagonal ensemble for chains with $L\approx 40$. 

%
\begin{figure}[t]
\begin{center}
\includegraphics*[width=0.87\linewidth]{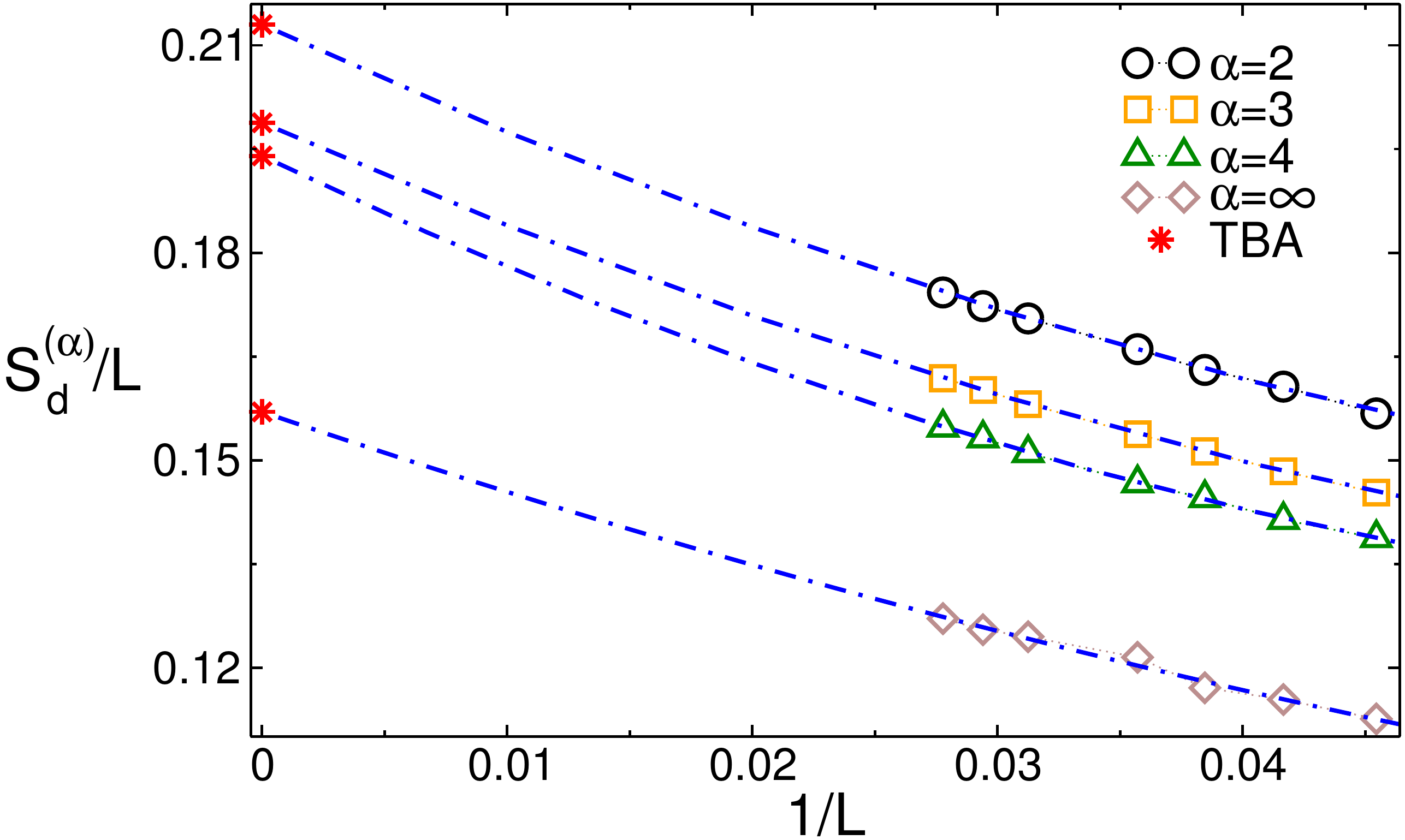}
\end{center}
\caption{ The R\'enyi diagonal entropies in the $XXX$ chain after the N\'eel quench. 
 The data are obtained by using the exact overlaps between the Bethe eigenstates and the N\'eel state. 
 The entropies density  $S_d^{(\alpha)}/L$ is plotted against the inverse chain length $1/L$ for several values of $\alpha$. 
 The star symbols are the Bethe ansatz results in the thermodynamic limit. 
 The dash-dotted lines are fits to $S_d^{ (\alpha)}=s_\infty^{(\alpha)} +a_\alpha/L+b_\alpha/L^2$, with $s_{\infty}^{(\alpha)}$ fixed to 
 the TBA result and $a_\alpha,b_\alpha$ fitting parameters. 
}
\label{bethe_f}
\end{figure}
%

Our exact diagonalisation results are  discussed in Figure~\ref{diag-ed}, showing 
$S_d^{(\alpha)}$ for $\alpha=2$, $\alpha=3$, $\alpha=4$ (panel (a), (b), and (c) respectively)  plotted versus $1/L$. 
Finite-size effects are visible for all values of $\alpha$. Interestingly, in the region $\Delta\approx 1$ these oscillate with the parity of $L/2$. 
The star symbols denote the diagonal entropy densities $s_{\infty}^{(\alpha)}$ in the thermodynamic limit,  
which are obtained using~\eref{d-renyi-main}. The dash-dotted lines are linear fits to $S_d^{(\alpha)}=s^{(\alpha)}_{\infty} + a_\alpha/L$, with 
$s^{(\alpha)}_\infty$ fixed by~\eref{d-renyi-main}. The agreement between the data and~\eref{d-renyi-main} is satisfactory 
for $\Delta=1$ and at large $\Delta$. For intermediate values of $\Delta$ the large oscillations do 
not allow for a reliable confirmation of the theoretical results, although the data 
are clearly compatible with~\eref{d-renyi-main}.

\subsubsection{Numerical Bethe ansatz,}
\label{sec:ba-num}

We now provide a further check of~\eref{d-renyi-main} by constructing the diagonal R\'enyi entropies 
using the exact overlaps between the eigenstates of the $XXZ$ model and the N\'eel state. These 
can be calculated from the solutions of the Bethe-Gaudin-Takahashi equations~\eref{bgt-eq} using 
the Algebraic Bethe Ansatz~\cite{pozsgay-14,bdwc-14,wdbf-14}. 
Although this is possible, in principle, for all the $XXZ$ chain eigenstates, 
a technical problem arises for eigenstates that correspond to solutions of the BGT equations 
containing zero momentum strings, i.e., with vanishing string center. Precisely, some fictitious 
singularities appear in the overlap formulas, which have to be removed. To overcome this issue one 
needs to go beyond the string hypothesis, considering the finite-size behaviour of the string deviations~\eref{str-hyp}. 
This is a formidable tasks that in practice can be performed only for small chains. 
It has been shown that in the thermodynamic limit the vast majority of 
eigenstates of the $XXZ$ chain with  finite N\'eel overlap contain zero-momentum strings. 
More precisely, the ratio between the total number of parity-invariant eigenstates $Z_{Neel}$ 
and the ones without zero-momentum strings $\widetilde{Z}_{Neel}$ vanishes in 
the thermodynamic limit as \cite{ac-16qa}
\begin{equation}
\label{r-neel}
\frac{\widetilde{Z}_{Neel}}{Z_{Neel}}\propto\frac{4}{\sqrt{\pi L}}. 
\end{equation}
As a consequence of~\eref{r-neel}, all the expectation values calculated on the restricted 
ensemble constructed by excluding the zero-momentum strings vanish in 
the thermodynamic limit. For instance, one has 
\begin{equation}
\label{weight}
w_{\Psi_0}\equiv \sum\limits_{m}'|\langle m|\Psi_0\rangle|^2\to 0, 
\end{equation}
where the prime is to stress that only eigenstates that do not contain zero-momentum strings are 
included in the sum. On the other hand, for any finite size and any normalised initial state $|\Psi_0\rangle$ 
it should be $\sum_m|\langle m|\Psi_0\rangle|^2=1$. 
However, it has been suggested in Ref.~\cite{wdbf-14} and confirmed numerically in~\cite{ac-16qa} that eigenstates containing 
zero-momentum strings are irrelevant in the thermodynamic limit. 
The idea proposed in~\cite{ac-16qa} is that the diagonal ensemble expectation values must be renormalised by the factor 
$w_{\Psi_0}$ in~\eref{weight}.
The diagonal R\'enyi entropies in this approach are given by 
\begin{equation}
\label{ba-ren}
S^{(\alpha)}_d=\frac{1}{1-\alpha}\ln\Big[\frac{1}{w_{\Psi_0}^{\alpha}}\sum\limits_{m}' |\langle m|\Psi_0\rangle|^{2\alpha}\Big].
\end{equation}
This reweighted expression converges for large $L$ to the thermodynamic expectation value, but has different $1/L$ 
finite-size corrections~\cite{ac-16qa}.

In Figure~\ref{bethe_f} we report the numerical results from \eref{d-renyi-main} for the N\'eel quench at $\Delta=1$. 
The symbols correspond to different values of $\alpha$ ranging from $\alpha=2$ to $\alpha=\infty$. 
In the Figure we plot $S^{(\alpha)}/L$ versus $1/L$ for chains with $L\le 38$. 
The star symbols are the theoretical results obtained using the $TBA$ approach (cf.~\eref{d-renyi-main}). 
The dash-dotted lines are fits to $S_d^{(\alpha)}/L=s_\infty^{(\alpha)}+a_\alpha/L+b_\alpha/L^2$, with 
$a_\alpha,b_\alpha$ fitting parameters and $s_\infty^{(\alpha)}$ the entropy density obtained from Bethe ansatz. 
The numerical results are clearly compatible with~\eref{d-renyi-main} in the thermodynamic limit.

\section{Entanglement versus diagonal entropies}
\label{sec:ent-vs-dia}

%
\begin{figure}[t]
\begin{center}
\includegraphics*[width=0.95\linewidth]{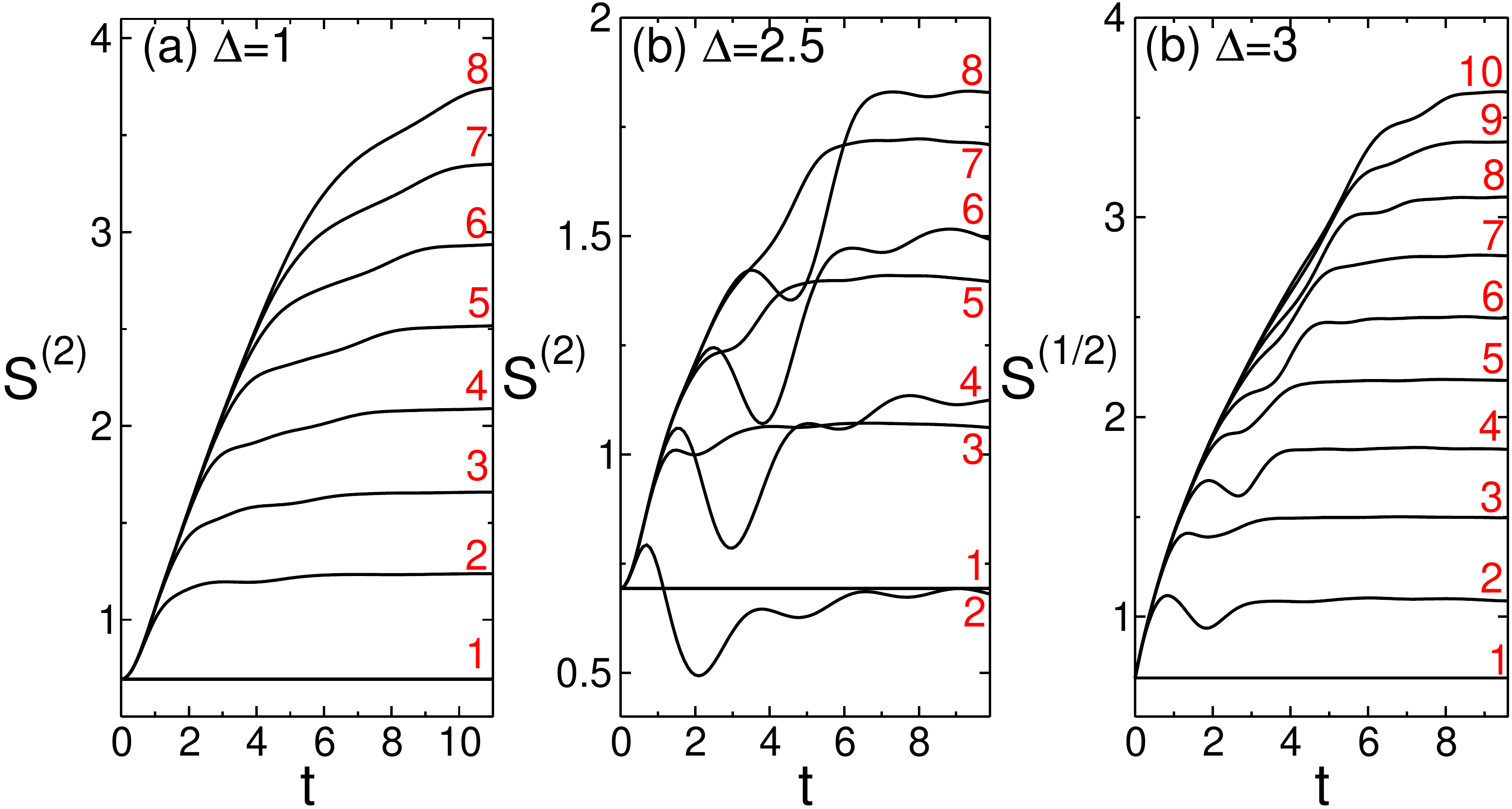}
\end{center}
\caption{R\'enyi entanglement entropies after the quench from the N\'eel state in the $XXZ$ chain: 
 tDMRG results for $L=40$  plotted as a function of time. 
 Different curves correspond to different subsystem sizes (accompanying numbers). 
 Panels (a) and (b) show data for $S^{(2)}$ at $\Delta=1$ and $\Delta=2.5$, respectively. 
 Panel (c) shows data for $S^{(1/2)}$ for $\Delta=3$. Note at short time the sizable oscillations for $\ell$ even. 
}
\label{renyi-dmrg1}
\end{figure}
%

In this section we discuss the stationary value of the R\'enyi entanglement entropies of a block $A$  of $\ell$ contiguous spins  
after the N\'eel quench in the $XXZ$ chain. 
As stressed in the introduction, the stationary value of the entanglement entropy is equal to the 
thermodynamic entropy (i.e. the entropy of the GGE) which is the double of the diagonal entropy \cite{ac-17}. 
This equivalence has been investigated in several studies for the von Neumann entropy, both for free 
systems \cite{gurarie-2013,collura-2014,fagotti-2013b,kormos-2014} and for interacting ones~\cite{dls-13,bam-15,piroli-2016,ac-17},
but not for the R\'enyi entropies with $\alpha\neq1$.
Here we then perform extensive tDMRG simulations~\cite{white-2004,daley-2004,uli-2005,uli-2011,itensor}, to provide numerical evidences that 
in the limit of large chains and large subsystems, i.e., $L,\ell\to\infty$ with $\ell\ll L$, one has 
\begin{equation}
\label{d-vn}
\frac{2S_d^{(\alpha)}}{L}=\frac{S^{(\alpha)}_{\rm GGE}}{L} =\frac{S^{(\alpha)}_{A}}{\ell}. 
\end{equation}
%


Before presenting our results for the R\'enyi entropies, it is worth to recall some generic features about the time evolution of the 
entanglement entropy after a quench. 
In many numerical and analytic calculations \cite{calabrese-2005,fagotti-2008,dmcf-06,ep-08,lauchli-2008,kim-2013,nr-14,coser-2014,cce-15,fc-15,chmm-16,buyskikh-2016,d-17,kctc-17,mkz-17,nahum-17},
as well as in one experiment \cite{kaufman-2016}, it has been observed that the entanglement entropy first grows linearly in time up to a time 
$t^*$ proportional to the length of the subsystem and subsequently slowly saturates to the extensive value of the thermodynamic ensemble. 
For an integrable model, this behaviour can be explained in terms of a quasi-particle picture introduced in \cite{calabrese-2005}. 
According to this picture, the prequench initial state acts as a source of pairs of quasiparticle excitations with velocity $v(\lambda)$. 
 Although quasiparticles created far apart from each other are incoherent, those emitted at the same point in space are entangled. 
 Because these propagate ballistically throughout the system, larger regions get entangled while time passes.
 At time $t$, the entanglement entropy is proportional to the total number of quasiparticle pairs that, emitted at the same point in space, 
 are shared between A and its complement.
When a maximum quasiparticle velocity $v_M$ exists, then for $t\leq \ell/(2v_M )=t^*$ the entropy increases linearly. 
This picture has been used in \cite{alba-2016} to provide a prediction for the entanglement entropy which becomes 
exact in the space-time scaling limit (i.e. for $t,\ell\to\infty$ with $t/\ell$ fixed). 
Conversely, for non-integrable models, quasiparticles have usually a finite life-time and the picture above could be used only to have some 
gross features for the time evolution of the entanglement entropy. 
However, the linear increase of the entanglement followed by saturation has been observed generically for non-integrable models 
and its origin is likely to have also an alternative explanation, see e.g. \cite{nahum-17,bhy-17}.

An overview of our tDMRG data for the entanglement entropy $S^{(\alpha)}_A$ is reported in Figure~\ref{renyi-dmrg1}. The data 
are for $\alpha=2$ (panel (a) and (b)) and $\alpha=1/2$ (panel (c)). In each panel, the different 
curves correspond to different subsystem sizes $\ell\lesssim 10$. The data are obtained by Trotter 
evolution of the Matrix Product State representation of the N\'eel state. The largest bond dimension 
employed in the simulation is $\chi=400$. The Trotter time discretisation step is $\delta t=0.05$. 
For $\Delta=1$, $S^{(2)}$ exhibits a quite linear smooth increase with time and a saturation at $t\propto\ell$, as it should. 
At larger $\Delta$, sizable odd-even effects are present (see (b) in the Figure) and the data for even $\ell$ show large oscillating corrections 
with time. This is likely to be imputed to the relative small value of $\ell$ accessible by tDMRG and these oscillations are expected to 
disappear in the space-time scaling limit.
A similar, but less pronounced, behaviour is observed for $\alpha=1/2$ ((c) in the Figure). 

\begin{figure}[t]
\begin{center}
\includegraphics*[width=0.88\linewidth]{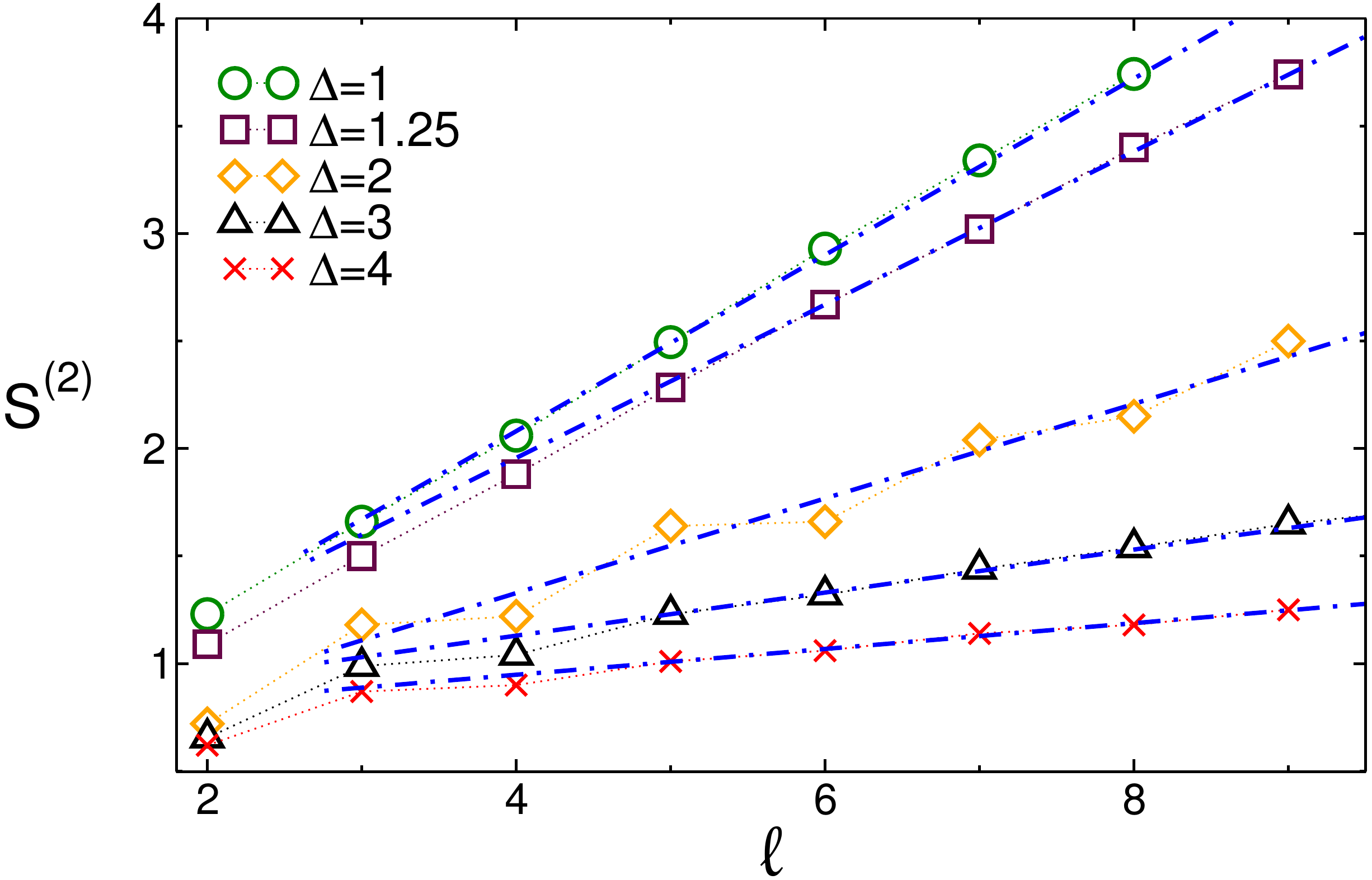}
\end{center}
\caption{Stationary R\'enyi entanglement entropy $S^{(2)}$ in the $XXZ$ chain for different values of  $\Delta$ as function the 
 subsystem length $\ell$. The points are tDMRG results for a chain with $L=40$ at $t\sim 10$. 
 The dash-dotted lines are fits to $S^{(2)}= s^{(2)}_{ \infty}\ell+a_2$, with $a_2$ a fitting parameter and $s^{(2)}_{\infty}$ fixed to 
 the density of the GGE R\'enyi entropy obtained from the Bethe ansatz. 
}
\label{renyi-dmrg2}
\end{figure}

In Figure~\ref{renyi-dmrg2} we focus on the steady-state value of $S^{(2)}$. 
The symbols denote the stationary values of $S^{(2)}$ (tDMRG data at $t\approx 10$, see Figure~\ref{renyi-dmrg1}) for different values of 
$\Delta$ plotted against the subsystem length $\ell$. The expected volume law $S^{(2)}\propto\ell$ is clearly visible. 
Moreover, $S^{(2)}$ decreases monotonically with increasing $\Delta$, similar to the von Neumann entropy~\cite{alba-2016}. 
This reflects that for $\Delta\to\infty$ the N\'eel state becomes the ground state of the $XXZ$ chain. 
The dash-dotted lines are linear fits to 
\begin{equation}
\label{fit}
S^{(\alpha)}=a_\alpha+s_\infty^{(\alpha)}\ell, 
\end{equation}
with $\alpha=2$, $a_2$ a fitting parameter and $s_\infty^{(2)}$ the density of the GGE R\'enyi entropy, 
as obtained from Bethe ansatz~\eref{d-renyi-main}. 
The agreement with the numerical data is perfect already for $\ell\gtrsim 5$ (although there are oscillations with the parity of $\ell$ for $\Delta=2$).  
This allows us to conclude that, within the system sizes accessible with DMRG, the numerical data confirm the validity of~\eref{d-vn}. 
%
\begin{figure}[t]
\begin{center}
\includegraphics*[width=0.88\linewidth]{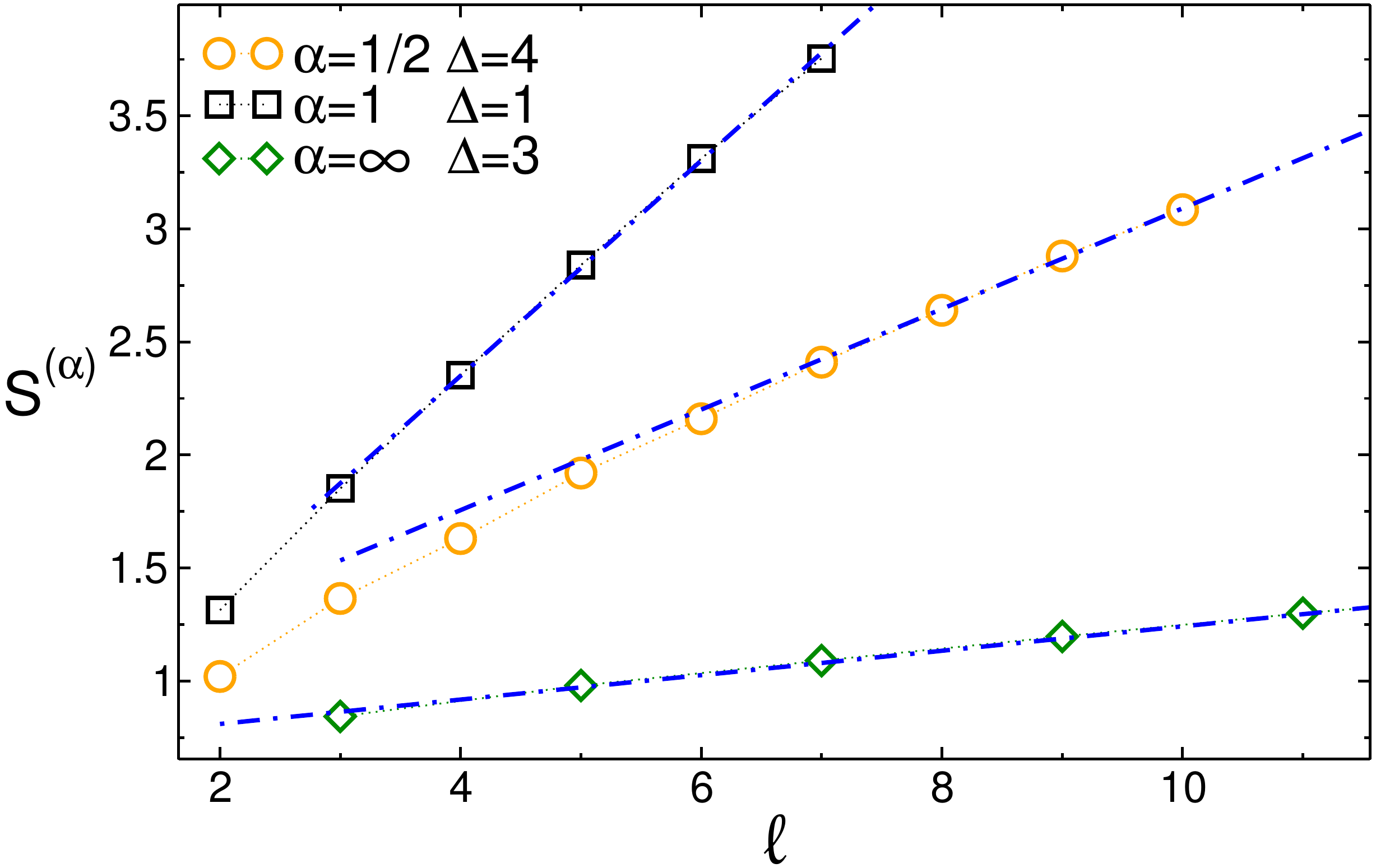}
\end{center}
\caption{Stationary R\'enyi entanglement entropies after the N\'eel quench  in the $XXZ$ chain for $\alpha=1/2,1,\infty$. 
The dash-dotted lines are linear fits to $S^{(\alpha)}= s_\infty^{(\alpha)}\ell+a_\alpha$, with $s_\infty^{(\alpha)}$ the Bethe ansatz 
 density of the GGE R\'enyi entropies and $a_\alpha$ a fitting parameter. 
}
\label{renyi-dmrg3}
\end{figure}
%
In Figure~\ref{renyi-dmrg3} we also consider other values of $\alpha$, namely 
$\alpha=1/2$ (circles) and $\alpha=\infty$ (diamonds) (also old data \cite{alba-2016} for $\alpha=1$ (squares) are shown for comparison). 
We restrict ourselves to the relatively large values of $\Delta=3,4$, for which longer times and larges subsystem sizes can be accessed with 
tDMRG, because of the mild entanglement increase after the quench. 
For $\alpha=\infty$ we only show  data for odd $\ell$ because for even $\ell$ severe finite-size corrections do not 
allow us to reliably extract the stationary value. 
Irrespective of $\alpha$, the entropies exhibit the expected volume-law behaviour at large $\ell$. 
The dash-dotted lines are fits to~\eref{fit}, with $a_\alpha$ a fitting parameter and $s_\infty^{(\alpha)}$ the GGE entropy density 
obtained using Bethe ansatz. 
For $\alpha=1/2$, large finite-size effects are present, and the data start becoming 
compatible with the asymptotic behaviour~\eref{fit} only for $\ell\gtrsim 7$. 
On the other hand, for $\alpha=\infty$, the data perfectly agree with~\eref{fit} already for $\ell\ge 5$.

\section{Conclusions} 
\label{sec:concl}

We presented a systematic study of the R\'enyi entropy after a quantum quench in the XXZ spin-chain starting from the N\'eel state. 
We employed a recently developed variation \cite{ac-17} of the quench action method \cite{caux-2013,caux-16} 
which provides the diagonal and GGE R\'enyi entropies as generalised free energy on a saddle point macrostate which is different from the one 
for local observables and von Neumann entropy. 
As first step we wrote explicit TBA equations \eref{qa-tba} (or equivalently  \eref{dec}) for the root densities describing the macrostate.
We did not manage to solve these equations analytically (while for $\alpha=1$ it is possible \cite{wdbf-14}) and so we mainly based our analysis 
on their exact numerical solution.  
Plugging these solutions in the saddle-point expectation \eref{r-theo-1}, we readily obtain the R\'enyi entropies for arbitrary order $\alpha$
and anisotropy $\Delta\geq1$.
An interesting first observation is that the integrated functions for the R\'enyi entropy ($(-\alpha\epsilon_n+s_n)$  in~\eref{r-theo-1}) 
are not positive for all $\lambda$, although $S_d^{(\alpha)}>0$.
Thus these quantities cannot be interpreted as the contribution of the quasiparticle $n$ of momentum $\lambda$ to the entropy, 
which must be positive. 
However, one can think of adding to~\eref{r-theo-1} some functions $r_n(\lambda)$ such that $\sum\int_0^{\pi/2} d\lambda r_n(\lambda)=0$; 
this addition does not change the result for $S_d^{(\alpha)}$, but alters the densities. 


There are few limits in which the TBA equations can be solved analytically. These serve as reference points for the numerical solutions 
and they provide very important insights about the overall structure of the solutions themselves.
The first limit we consider is the one for large $\Delta$ which provides a systematic expansion in powers of $\Delta^{-1}$ 
for arbitrary $\alpha$. 
For $\Delta=\infty$ all entropies tend to zero, reflecting the fact that for $\Delta=\infty$ the N\'eel state is the ground-state of the XXZ chain. 
The other limits in which we work out the entire solution correspond to specific values of $\alpha=0,1,\infty$ (which are the max, the 
von Neumann, and min entropy respectively). 
While the results for $\alpha=0$ and $\alpha=1$ have been known by other means and only represent consistency checks for the 
general approach, the results for $\alpha=\infty$ are new and insightful. 
From the definition~\eref{d-renyi}, the min diagonal entropy is determined by the eigenstate with the largest N\'eel overlap. 
For $\Delta>\Delta^*$ this is the ground state of the $XXZ$ chain, while for $\Delta<\Delta^*$
the min entropy is determined by a finite energy density excited state that we calculate. 
The transition between these two regimes happens at a special value of anisotropy $\Delta^*\approx 1.76692$.
In both regimes the min entropy is determined by a state with zero Yang-Yang entropy. 
However, an important difference is that at small $\Delta$ the macrostate contains bound-states with an arbitrary number of particles, 
but for $\Delta>\Delta^*$ only one-strings are present. 
We do not expect the relation between the min entropy and the ground state of the $XXZ$ model to be true for other initial states, 
because this follows  from N\'eel state being the ground-state for large $\Delta$.
Conversely,  the fact that the min entropy is determined by a state with zero Yang-Yang entropy might be generic. 
This aspect deserves further investigation and could remain true in non-integrable models.

We numerically test our results by constructing explicitly the diagonal ensemble for finite-size $XXZ$ chains. 
We use both exact (full) diagonalisation and a numerical Monte Carlo implementation \cite{ac-16qa} of the Bethe ansatz based on the 
exact formulas~\cite{pozsgay-14,wdbf-14} for the overlaps with the N\'eel state.  
Extrapolating the results to the thermodynamic limit (carefully accounting for finite size effects), we find that the numerical data perfectly match 
the TBA predictions.  

We investigated the relation between the diagonal/GGE entropies and the entanglement R\'enyi entropies. 
The latter are obtained using time-dependent Density Matrix Renormalisation Group simulations. 
Our results confirm that for any $\alpha$ the entanglement R\'enyi entropy density is compatible with  
the density of the thermodynamic entropy obtained by TBA.
This result however is not sufficient to permit the reconstruction of the full-time dynamics 
of the R\'enyi entropies, as done for the von Neumann entropy~\cite{alba-2016},
 by applying the semiclassical picture of Ref.~\cite{calabrese-2005}.
The bottleneck in  this reasoning is that the thermodynamic entropies are not written in terms of 
the saddle point root densities describing local observables and 
only the latter densities correspond to the quasiparticles with a semiclassical dynamics.

It is highly desirable to extend our analysis to quenches from different initial states that can be solved by means of quench action and  
Bethe ansatz, such as the Majumdar-Ghosh state, the tilted N\'eel state, and the tilted ferromagnet. 
The extension to these states should allow to clarify which aspects found here are general and which ones are instead specific of the N\'eel quench.

Another interesting research direction would be to use the results for the R\'enyi entanglement entropies to derive the distribution 
of the entanglement spectrum levels in the steady state, as it has been done at equilibrium in Refs.~\cite{calabrese-2008,act-17}. 
In contrast with equilibrium, the structure of the entanglement spectrum after quantum quenches has not been investigated in 
detail, although some results are already available~\cite{poilblanc-2011,torlai-2014,canovi-2014,jhu-17}. 
This calculation requires the knowledge of the analytical dependence on $\alpha$ of the R\'enyi entropies, which, although it is not expected to 
be simple for out-of-equilibrium systems, it should be obtainable in some limits, for instance at large $\Delta$ and for some specific initial 
states.

\section{Acknowledgments} 
VA acknowledges support from the European Union's Horizon 2020 under the Marie Sklodowska-Curie grant agreement No 702612 OEMBS.

\appendix

\section{The diagonal entropy is half of the Yang-Yang entropy} 
\label{app1}

In Ref. \cite{ac-17} it has been shown in very general terms that the diagonal entropy is half of the Yang-Yang entropy, i.e. 
that in the limit $\alpha\to 1$ 
\begin{equation}
\label{limit}
\lim_{\alpha\to 1}S^{(\alpha)}_d=S_d\equiv -{\rm Tr} \rho_d\ln\rho_d=-\frac12 S_{YY}. 
\end{equation}
It is not so straightforward to recover this results from the overlap TBA equation, but it is possible, as we show in this appendix.

The strategy is to consider small deviations of the root densities $\rho_n$ and $\eta_n$ around their values at $\alpha=1$. 
For $\alpha\to 1$, we can  write $\eta_n$ as
\begin{equation}
\label{etans}
\eta_n=\eta^{(0)}_n+\eta_n',\quad\textrm{with}\,\eta'_n\ll\eta_n^{(0)}, 
\end{equation}
where $\eta_n^{(0)}$ are the solutions of the TBA equations for $\alpha=1$ and 
$\eta'_n$ is a ${\mathcal O}(\alpha-1)$ correction. 
Plugging \eref{etans} into~\eref{dec}, and keeping linear terms in $\eta_n'$, one obtains 
the infinite system of equations 
\begin{equation}
\label{a1}
\frac{\eta_n'}{\eta^{(0)}_n}=(\alpha-1) d_n+s\star\Big[\frac{\eta_{n-1}'}
{1+\eta_{n-1}^{(0)}}+\frac{\eta'_{n+1}}{1+\eta^{(0)}_{n+1}}\Big]. 
\end{equation}
We now move to the densities $\rho_n$ and we use the ansatz 
\begin{equation}
\label{rhons}
\rho_n=\rho^{(0)}_n+\rho_n'\quad\textrm{with}\,\rho_n'\ll\rho_n^{(0)}. 
\end{equation}
Plugging~\eref{rhons} in~\eref{tba-eq} and keeping the leading order in $\rho'$, one obtains 
\begin{equation}
\label{rho'}
\rho_n^{(0)}\eta_n'+\rho_n'(1+\eta_n^{(0)})=
-\sum\limits_{m}a_{n,m}\star\rho_m'. 
\end{equation}
In deriving~\eref{rho'} we used that the constraint on the magnetisation $\sum_m m\int d\lambda\rho_m=1/2$ implies
\begin{equation}
\sum\limits_{m}m\int_{-\pi/2}^{\pi/2}d\lambda\rho'_m(\lambda)=0, 
\end{equation}
because $\sum_m m\int d\lambda\rho_m^{(0)}=1/2$. 
This allows to neglect the term with the magnetic field in~\eref{tba-eq}. 

We are finally ready to consider the diagonal entropies \eref{r-res}, writing  for $\alpha\to 1$ 
\bea
{\mathcal E}&=&{\mathcal E}^{(0)}+{\mathcal E}'\\
S_{YY}&=&S_{YY}^{(0)}+S_{YY}', 
\eea
where ${\cal E}'$ and $S_{YY}'$ are ${\mathcal O}(\alpha-1)$. Plugging in the definitions 
of ${\cal E}$ and $S_{YY}$ Eqs. \eref{etans} and\eref{rhons}, it is straightforward to 
derive that 
\bea
\label{ep}
 {\cal E}'&=&\frac{L}{2}\sum_n\int_0^{\pi/2}d\lambda\rho_n'(\lambda)g_n,\\\label{sp}
 S_{YY}'&=&L\sum_n\int_{-\pi/2}^{\pi/2}d\lambda[(\rho_n\eta'_n+\eta_n\rho_n')\ln(1+\eta_n^{-1})
+\rho_n'\ln(1+\eta_n)]. 
\eea

Let us now multiply~\eref{rho'} by $\ln(1+\eta_n^{-1})$, take the sum over $n$ and integrate over the rapidity, to obtain
\bea\fl
\label{c1}
\sum_n\int_{0}^{\pi/2}d\lambda[(\rho_n\eta'_n+\eta_n\rho_n')\ln(1+\eta_n^{-1})+
\rho'_n\ln(1+\eta_n^{-1})] \\
=\sum_{m,n}\int_0^{\pi/2} d\lambda d\mu\ln(1+\eta_n^{-1}
(\lambda))(a_{nm}(\lambda-\mu)+a_{nm}(\lambda+\mu))\rho'_m(\mu).\nonumber 
\eea
Similarly, multiplying~\eref{qa-tba} by $\rho_n'$ for $\alpha=1$, summing 
over $n$ and integrating over $\lambda$, one obtains 
\bea\fl
\label{c2}
\sum_n\int_{0}^{\pi/2}d\lambda[\ln(\eta_n)-g_n]\rho_n'(\lambda)=\\
=-\sum_{m,n}\int_0^{\pi/2} d\lambda d\mu\ln(1+\eta_n^{-1}(\mu))
(a_{nm}(\lambda-\mu)+a_{nm}(\lambda+\mu))\rho'_m(\lambda).
\eea
The right-hand-side of~\eref{c2},  coincides with minus the right-hand side in~\eref{c1}, because $a_{nm}(\lambda-\mu)=
a_{nm}(\mu-\lambda)$. Thus summing~\eref{c1} and~\eref{c2} we have 
\begin{equation}
\label{tri}
-2{\cal E}'+\frac{1}{2}S_{YY}'=0.
\end{equation}

We finally have
\be\fl
\label{prove}
-2\alpha{\cal E}+\frac{1}{2}S_{YY}\approx \\-2{\cal E}^{(0)}+\frac{1}{2}S_{YY}^{(0)}
-2{\cal E}'+\frac{1}{2}S_{YY}'-2(\alpha-1){\cal E}^{(0)}=
-2(\alpha-1){\cal E}^{(0)},
\ee
where in the last step we used~\eref{tri} and~\eref{tri-0}. 
This is equivalent to
\begin{equation}
\lim_{\alpha\to 1}S^{(\alpha)}_d=S_d=\frac12 S_{YY}
\end{equation}
A fundamental consequence of this equation is  that the diagonal entropy $S_d$ 
is determined by the saddle point at $\alpha=1$, which describes local and 
quasilocal observables in the steady state after the quench. 
This is just a consequence of the fact that the logarithm
in the definition of $S_d$ \eref{limit} cannot shift the saddle point of the TBA.




\end{document}